\RequirePackage{fix-cm}
\documentclass[twocolumn,epjc3]{svjour3}  
\usepackage{tabularx}
\usepackage{orcidlink}
\usepackage{caption}
\usepackage{soul}
\usepackage{xcolor}
\usepackage{booktabs}
\usepackage{lineno}
\usepackage{amsmath}
\usepackage{bm}
\usepackage[labelfont=bf]{caption}
\renewcommand{\figurename}{Fig.}

\smartqed  
\RequirePackage{graphicx}
\journalname{Eur. Phys. J. C}
\begin{document}

\title{Particle tracking at high luminosities using a novel reconstruction approach}

\author{B. K. Sirasva\hspace{-0.02em}\thanksref{addr1}\orcidlink{0009-0009-8508-9641}\and S. Dugad\hspace{-0.02em}\thanksref{e1,addr1, addr5}\orcidlink{0009-0008-1407-5296}\and Y. Kumar\hspace{-0.02em}\thanksref{addr1}\orcidlink{0009-0001-7980-4305}\and P. Suryadevra\hspace{-0.02em}\thanksref{addr3}\orcidlink{0000-0002-5381-7570}\and M. Yadav\hspace{-0.02em}\thanksref{addr2}\orcidlink{0009-0008-2765-6751}
}

\thankstext{e1}{e-mail: shashikant.dugad@cern.ch}


\institute{Department of Physical Sciences, Indian Institute of Science Education and Research Mohali, Punjab, India 140306 \label{addr1}
            \and
            Department of Computer Engineering, Fr. C. Rodrigues Institute of Technology, Vashi, Maharashtra, India 400703 \label{addr5}
           \and
           Department of High Energy Physics, Tata Institute of Fundamental Research, Mumbai, Maharashtra, India 400005 \label{addr3}
           \and
           School of Computer Engineering, Kalinga Institute of Industrial Technology, Bhubaneswar, Odisha, India, 751024\label{addr2}      
}

\date{}

\sloppy
\maketitle

\begin{abstract}

Tracking charged particles with high precision is of vital importance for collider experiment like those operating at the Large Hadron Collider (LHC), CERN. The tracking detector in the CMS experiment is composed of multi-layer silicon based tracker with 3-dimensional position sensitivity. High precision position data from tracker operated in high magnetic field, is used to reconstruct the trajectories of charged particles and obtain their kinematic parameters ($p_t, \eta_0, \phi_0$) with high accuracy. In this paper, for Phase2 CMS tracker design, we present a novel track reconstruction algorithm for high luminosity (HL) era of the LHC. The algorithm identifies hits associated with each track and utilizes them to accurately determine the kinematic parameters of each track using machine learning (ML) architecture. The proposed algorithm has been applied on a large
sample of hard interactions simulated at high luminosity (HL) era of the LHC using \textsc{Pythia8} and \textsc{Geant4} framework for equivalent geometry of the outer tracker of the CMS experiment. Performance of the proposed algorithm has been studied using the key indicators such as reconstruction efficiency, fake rate and resolution. Comparison with traditional methods demonstrates robust performance with excellent efficiency and resolution with minimal fake rate.

\end{abstract}

\section{Introduction}\label{sec:intro}
The Large Hadron Collider (LHC) ~\cite{Lyndon_Evans_2008} at CERN is the world's most powerful particle accelerator. It accelerates high intensity bunches of protons up to 7 TeV in a 27-km underground ring in a clockwise and counter clockwise directions with the collision frequency of 40 MHz. The counter-rotating proton bunches collide at four interaction points, each hosting a major experiment: ATLAS ~\cite{The_ATLAS_Collaboration_2008}, CMS ~\cite{The_CMS_Collaboration_2008}, ALICE ~\cite{The_ALICE_Collaboration_2008}, and LHCb ~\cite{Alves:1129809}. State of art detectors deployed around the collision points are designed to address some of the fundamental questions about the evolution of the universe. The Compact Muon Solenoid (CMS) ~\cite{The_CMS_Collaboration_2008} and ATLAS ~\cite{The_ATLAS_Collaboration_2008} experiments at LHC are mainly designed to probe predictions of the Standard Model (SM) and beyond SM. A major milestone achieved by these detectors was the discovery of the Higgs boson \cite{ATLAS:2012yve}, confirming the final missing component of the SM. The high luminosity era of LHC (HL-LHC) provides a unique opportunity to precisely measure Higgs boson couplings and decay properties, which may reveal signs of new physics \cite{PhysRevD.110.032012}. 

The LHC is set to undergo a major upgrade to significantly enhance the instantaneous luminosity of the proton beam from $2 \times 10^{34} cm^2/sec^{-1}$ to $5-7.5 \times 10^{34} cm^2/sec^{-1}$ with a possibility of peak luminosity going up to $2 \times 10^{35} cm^2/sec^{-1}$ in the beginning of the fill ~\cite{ZurbanoFernandez:2020cco}. The CMS and ATLAS experiments are expected to acquire collision data with an integrated luminosity of $\sim 3000-4000 fb^{-1}$ over a period of 10 years of high luminosity operation of LHC. Significant increase in the integrated luminosity will enable precision measurements of Higgs properties,  discovery potential beyond the SM with the higher sensitivity. However, a substantial enhancement in the instantaneous luminosity will lead to larger number of simultaneous pile-up interactions per bunch crossing, posing a major challenge for the experiments because of the performance degradation of the detector due to higher integrated radiation dose. Hence, such an environment demands an upgrade of various subsystems of all the LHC detectors in order to withstand higher radiation dose and provide higher data rate handling capability during the entire operational period of HL-LHC. The upgrade phase of the CMS detector for HL-LHC is referred to as the {\it Phase2 upgrade} ~\cite{CERN-LHCC-2017-009}.

The CMS detector, designed to measure kinematic parameters of various particles, is mainly composed of a) tracking system b) electromagnetic and hadron calorimeters and c) muon system. The largest super conducting solenoid magnet with a magnetic field of 3.8 Tesla along the beam-axis, encloses tracker and calorimeter systems whereas, the outermost muon system is enclosed in a return magnetic field directed by the yoke sandwiched between successive layers of muon detector. Silicon based tracker detector accurately reconstructs the trajectory of each charged particle, whereas, calorimeters measure energy and direction of electrons, photons, and hadronic jets. Particles going out of the calorimeter are likely to be muons and their kinematic parameters are measured by sampling its trajectory by muon detector operating in presence of return magnetic field. Reconstructing the entire event by integrating data from all sub-systems of the CMS provides a comprehensive physical nature of the collision event ~\cite{Sirunyan_2017}. 

In this paper, we propose a novel approach to reconstruct particle trajectories in the HL-LHC era  ~\cite{Apollinari:2284929}, demonstrated using the simulated data from the CMS outer tracking system. The CMS Phase 2 upgrade entails a complete overhaul of the tracking system, the salient capabilities of the new tracker system are described below. The Phase-2 Tracker will have pseudorapidity coverage of $|\eta|<4$ with a higher granularity and higher readout bandwidth in order to sustain a high pile-up environment. It is designed to sample trajectories of a large number of charged particles produced by an extremely high number of simultaneous pp collisions (up to 200) in each bunch crossing. Basic detection elements used to build the tracker ~\cite{CERN-LHCC-2017-009} are essentially {\it micro-pixel} (PX), {\it macro-pixel strip} (PS), and {\it two strip} (2S) silicon sensor modules. Positional resolutions provided by each detection element is a) $25 \ \mu m \times 100 \ \mu m$ and $50 \ \mu m \times 50 \ \mu m$ for PX modules, b) $100 \ \mu m \times 1.5 \ mm$ for macro pixel and $100 \ \mu m \times 2.4 \ cm$ for PS modules and c) $90 \ \mu m \times 5.0 \ cm$ for 2S modules along $r-\phi$ and z (beamline) axis respectively. The tracker system consists of two main components, namely, {\it Inner} and {\it Outer} Tracker system. 

{\it Inner Tracker} (IT) system is located closest to the beamline and built with PX modules providing finest position resolution in all directions. The IT system is built with three different substructures: a) the barrel part (TBPX) made of four concentric layers; b) the forward part (TFPX) made of eight small disks followed by c) the endcap part (TEPX) having 4 large disks. All the 12 disks are mounted vertically on both sides of the interaction point in symmetrical manner. The {\it Outer Tracker} (OT) system consists of three parts: a) inner barrel part composed of three layers with strips and macro-pixel sensors modules (TBPS), b) outer barrel part consisting of three outer layers with two strips sensor modules (TB2S) and c) endcap (TEDD) made of five double disks with both PS and 2S modules on each side of the interaction pint~\cite{CERN-LHCC-2017-009}. 

\begin{figure}[!h]
    \centering
    \includegraphics[width=0.95\linewidth]{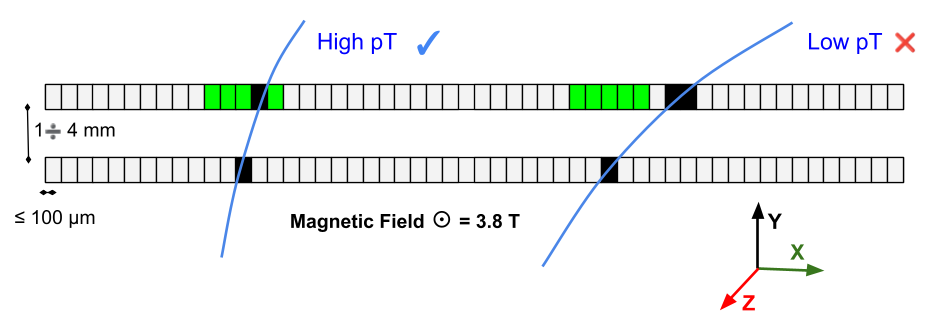}
    \caption{Schematic of signal correlation in nearby sensor layer. Green channels show the window used to select accepted stubs and reject low $p_t$ particles ($p_t<2 \ \rm GeV/c$).}
    \label{fig:stub}
\end{figure}

The CMS Phase-2 outer tracker system implements hit selection at the Level-1 (L1) trigger itself in order to have efficient trigger performance in the high pileup environment. Since transferring the full detector data at the 40 MHz bunch crossing rate is limited by the data bandwidth, the L1 trigger system of outer tracker identifies and eliminates hits associated with low transverse momentum ($p_{t}$) particles, directly at the front end. This is achieved by using two closely spaced silicon sensors, separated by 1.8 mm to 4 mm, within the same module. The low $p_t$ tracks experience larger curvature in the 3.8 T magnetic field resulting in higher azimuthal displacements between the hits registered in a closely spaced sensors as shown in Fig.~\ref{fig:stub}. By applying a programmable correlation window, the hits due to low $p_t$ particles and uncorrelated noise can be rejected. Accepted correlated hits are called {\it stubs}, with a typical programmable stub-generation threshold kept at  2 GeV/c. 

We begin by presenting overview of existing track reconstruction algorithms used in high-luminosity pp collisions in Section ~\ref{sec:trk_reco}. The work presented in this paper mainly uses the stubs with $p_t > 2 \ \rm GeV/c $ selected by the L1 trigger system of the OT for the reconstruction of tracks. In Section ~\ref{sec:TRF}, a novel filter parameter, $\alpha$, is introduced to efficiently associate {\it stubs} recorded by the OT system to their corresponding particle tracks. A toy-model tracker data is generated to demonstrate the potential of the proposed filter parameter $\alpha$ for mapping hits {\it directly} to their corresponding tracks. Large number of $t\bar t$ events with high pileup ($nPU=150 \pm 12.2$) are generated using \textsc{Pythia8} for more realistic validation of the proposed algorithm, described in Section ~\ref{sec:evt_gen}. Propagation of each particle from the vertex through a equivalent CMS outer tracker geometry is simulated using \textsc{Geant4} ~\cite{GEANT4:2002zbu}. Details of geometry construction using \textsc{Geant4}, geometry validation and tracklet formation is described in Section ~\ref{sec:geant4}.
 Reconstructed tracklets ($p_t > 2 \ \rm GeV/c $) are used for prediction of kinematic parameter  of tracks using ML-based approach presented in Section ~\ref{sec:MLT}.

\section{Overview of Track Reconstruction Methods}\label{sec:trk_reco}
Reconstruction of a large number of charged-particle trajectories in collider experiments is a computationally challenging task. Collective analysis of reconstructed kinematic parameters of each track, such as charge (q), transverse momentum ($p_t$), pseudo-rapidity ($\eta_0$), and azimuthal angle ($\phi_0$), provide crucial information about the underlying physics processes. Tracking algorithm is expected to deliver high track finding efficiency with low fake rate and good momentum and vertex resolution over a wide range of detector acceptance at high luminosities. Among the various approaches, the {\it Kalman Filter (KF)} ~\cite{Fruhwirth:1987fm} remains the most widely used tracking algorithm due to its robustness in handling detector resolution, multiple scattering, and energy-loss effects during track fitting. The KF iteratively estimates track parameters as particles propagate through successive detector layers. In recent years, Particle Flow (PF) (\cite{Sirunyan_2017}, \cite{mokhtar2026progress}) and machine learning (ML) based techniques have emerged as promising alternatives for track reconstruction. In the CMS Phase-2 upgrade, the KF, PF, and the ML-based reconstruction algorithms are being probed for estimating track parameters and their uncertainties as described below. 

\subsection{Kalman Filter (KF)}\label{sec:KF}
The CMS detector employs the Kalman Filter (KF) as one of its primary track reconstruction algorithms ~\cite{Speer:2005dp}. The KF is a recursive method that combines detector measurements with a physical propagation model to estimate track parameters, such as $p_t$, $\eta_0$, and $\phi_0$, in a magnetic field of 3.8 Tesla. Potential track candidates, referred as {\it tracklets}, consisting of stubs from each layer are formed using the {\it Combinatorial Track Finder} (CTF) approach recursively  ~\cite{elmetenawee2023cmstrackingperformancerun}. A pair of stubs in successive layers consistent with the beam spot position is fed as seed in the CTF. Tracklets having stubs in at least four OT layers are considered for the KF fitting.

The KF estimates the state vector, consisting of track parameters such as $q/p_t$, $d_0$ (transverse impact parameter), $z_0$, $\theta_0$ (angle {\it w.r.t.} beam axis), and $\phi_0$, along with their associated covariance matrix. At each tracker layer, the algorithm performs two sequential steps: prediction and update. In the prediction step, the state vector and covariance matrix are propagated to the next detector layer using the track propagation model. In the update step, the predicted state is refined using the measured stub positions and their uncertainties. The relative uncertainty between the predicted and measured quantities determines the {\it Kalman gain}, which governs the contribution of the new measurement to the updated state estimate. Stubs inconsistent with the track hypothesis can be rejected during the filtering process, thereby reducing fake track candidates. A backward smoothing procedure is often applied after filtering to further improve the final track parameter estimation ~\cite{Fruhwirth:1987fm}. 

Owing to the high luminosity conditions at the LHC, the track reconstruction algorithms must be optimized to achieve large computational throughput. The computational performance of the KF is improved by partitioning detector hits into $\eta$--$\phi$ regions and vectorizing operations on small sized $(5\times5)$ ~\cite{Gagnon_2022} state transition matrices. The Kalman Filter (KF) remains the baseline approach, with continuous developments focused on improving computational efficiency using modern computing architectures ~\cite{Gagnon_2022}. Complexity of the CTF coupled with KF scales exponentially with increasing pile up interactions, thereby, significantly increasing computational time, posing a major challenge at HL-LHC era of the LHC. 

\subsection{Particle Flow (PF) Algorithm} \label{sec:PF}

An improved global event reconstruction can be achieved by combining data from all the sub-detectors, such as tracker, calorimeter and muon detector to form tracks and clusters. Such an integrated approach is referred as particle-flow (PF) reconstruction and it is used to reconstruct the final state particles produced in pp collision (\cite{Sirunyan_2017}, \cite{mokhtar2026progress}). The kinematic parameters of charged hadrons are dominantly determined by the superior tracker resolution whereas the photons and neutral hadrons are in general identified by ECAL and HCAL clusters with no links in the tracker. Electrons are identified by a track in tracker with a major cluster in ECAL and significantly smaller trace in corresponding HCAL region. A momentum-to-energy ratio compatibility with the unity is used to improve resolution. Muons are identified by a track in the tracker with a corresponding link in the muon detectors. The PF concept was first developed by the ALEPH experiment at LEP ~\cite{DECAMP1990121}. The CMS experiment is among the first hadron collider experiments where the PF algorithm has been implemented successfully. It has been a focus of development in light of increased pileup and detector granularity for HL-LHC environment.

\subsection{Machine Learning (ML) based  Algorithms} \label{sec:ML_Track}
The $21^{\mathrm{st}}$ century has witnessed rapid advances in deep learning techniques ~\cite{LeCun:2015} with  applications in collider experiments, including data quality monitoring and reconstruction of complex events ~\cite{Radovic:2018dip}. There are several track reconstruction algorithms based on machine learning (ML) techniques. Collider data are often better described in the graphical form ~\cite{Bronstein_2017}. One of the most promising ML-based approaches for track reconstruction at the HL-LHC is based on {\it Graph Neural Networks} (GNNs) ~\cite{Shlomi_2021}. GNNs operate naturally on graph-structured data, where tracker hits are treated as nodes and potential track segments as edges. The network identifies edges compatible with charged-particle trajectories to construct complete track-hit associations. Although GNN training requires large datasets and substantial computational resources, efficient implementations on GPU and multi-core CPU architectures provide high computational throughput using parallel data processing ~\cite{Guiang:2882251,Brown:2875830}. 

The Hough Transform (HT) ~\cite{Hough_1962_2186,10.1145/361237.361242} is an image processing technique used to identify points lying on parametric curves, such as straight lines or circles, and is widely applied in pattern recognition. Although traditional implementations of HT ~\cite{ILLINGWORTH198887} are computationally expensive, several optimized versions ~\cite{James:2273410} have been developed for track finding in the collider experiments. Computational time of HT scales approximately linearly with the number of hits, hence, HT based approaches provides advantageous in the high-pileup conditions at the HL-LHC.

\begin{figure}[t]
    \centering
    \includegraphics[width = .48\textwidth]{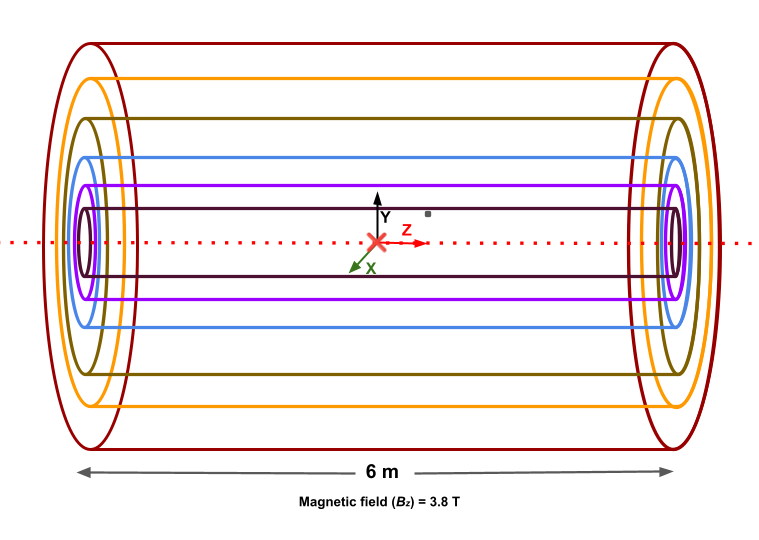}
    \caption{Schematic view of the six cylindrical silicon layers of the outer tacker.}
    \label{fig:tracker}
\end{figure}

\begin{table}[h]
  \centering
  \renewcommand{\arraystretch}{1.4} 
  \begin{tabularx}{\linewidth}{ccc}
    \toprule
    \textbf{Layer No.}   & \textbf {Radius (mm)}  & \textbf{Pixel Dimension} \\
    &  &  \textbf{$\textbf {r-z} \bm{\times} \textbf{r-}\bm{\phi}$} \\
    \midrule
        1  &  240 & $1.5 \ mm \times 90 \mu m$ \\
        2  &  360 & $1.5 \ mm \times 90 \mu m$ \\
        3  &  520 & $1.5 \ mm \times 90 \mu m$ \\
        4  &  690 & $50  \ mm \times 90 \mu m$ \\
        5  &  870 & $50  \ mm \times 90 \mu m$ \\
        6  & 1130 & $50  \ mm \times 90 \mu m$ \\
    \bottomrule
  \end{tabularx}
  \caption{Dimensions of prototype outer tracker built using \textsc{Geant4}. Length and thickness of each silicon cylinder is set at 6 meter and $300 \ \mu m$ respectively.}
  \label{tab:det_dimn}
\end{table}

\section{Tracker Simulation Framework (TSF)}\label{sec:TRF}

In this section, the filter parameter $\bf \alpha $ is introduced and its effectiveness in formulating tracklets, {\it i.e.}, efficiently and reliably mapping stubs to their respective particle trajectory is demonstrated. To begin with, the motion of a large number of charged particles traversing through the toy detector model of the outer tracker is simulated. A simplistic toy detector model is built by having six concentric cylinders made of silicon ($300\  \mu m$ thick), are symmetrically placed around the origin, with their axis aligning with the beam axis as shown in Figure \ref{fig:tracker}. Detector and pixel dimensions are chosen to be similar to the design parameters of the outer tracker geometry of the CMS experiment (Table \ref{tab:det_dimn}). It is to be noted that, since the width of pixels in the azimuthal direction is same ($90 \ \mu m$) for all layers, the azimuthal position ($\Delta\phi$) resolution improves with the increasing layer number. Particles with known momentum ($p_t >2 \ \rm GeV/c$) are projected {\it geometrically} into the outer tracker region in the presence of a longitudinal magnetic field of 3.8 Tesla. As explained before, in the CMS outer tracker, stubs are formed only for those particles having their transverse momentum above the threshold of $2 \ \rm GeV/c$. It is to be noted that all the charged particles satisfying the {\it stubs} requirement  leave the tracker without performing a helical trajectory in the tracker, which ensures the particle's trajectory intersects each of the tracker layers {\it only once}. These charged particles, after leaving the tracker loose energy either in the calorimeter ($e^\pm, \ \pi^\pm, \ K^\pm \ etc. $) or leave the CMS detector through the muon detector ($\mu^\pm$). 

The {\it Tracker Simulation Framework} (TSF) enables precise {\it geometrical projection} of charged particles trajectories from vertex into the tracker region with a known charge and momentum vector ($q, \ p_t, \ \phi_0, \ \rm and \ \theta_0$) in a uniform magnetic field of 3.8 Tesla along the beam axis. $\theta_0$ is the angle made with the beam axis. The TSF  provides intersection coordinates ($x_i^L, \ y_i^L, \ z_i^L$) of the trajectory with each of the tracker layers (L), using which, their pixel indices ($\phi_i^{L}, \ z_i^{L}$) are obtained for each layer. Large number of tracks with randomly chosen a) $\theta$ between $0-180^0$, b) $\phi_0$ between $0-360^0$ and c) $p_t$ between $2-p_t^{max} \ \rm GeV/c$ are propagated through the tracker geometry. Kinematic parameters and pixel indices of all tracks are stored in the database. These datasets in different ranges of $p_t$ are used to explore the potential of the {\it filter parameter} $\boldsymbol\alpha$ for efficient and robust mapping of stubs with their corresponding tracks. Same datasets are used to train ML models by keeping kinematic parameters as target class of the model. Trained model is used to predict kinematic parameters of reconstructed tracklets consisting of mapped stubs. Since the data is generated only through geometrical projection, it does not have any systematic effects due to energy loss, thus, provides a good handle to assess the best possible performance of the tracklet formation using filter parameters. 

\subsection{Filter Parameter $\boldsymbol{\alpha}$}\label{sec:fp}
In the CMS experiment, charged particle with transverse momentum $p_t$ performs circular trajectory in the transverse plane in presence of uniform longitudinal magnetic field of $B_z=3.8$ Tesla with a radius of curvature $R_p$ given by,

\begin{equation}\label{eqn:radius}
 R_p \ (meter) =  \frac{p_t \ (GeV/c)} {0.3 \times q \times B_z \ (Tesla)} 
\end{equation}

It can be seen from above equation (Eqn. \ref{eqn:radius}), for $B_z = 3.8 \ \rm Tesla$, particles with $p_t>0.7 \ \rm GeV/c$ has a diameter of curvature more than radius of the outermost layer of the tracker ($R_6 = 1.11$ meter). Hence, particles with $p_t>0.7 \ \rm GeV/c$ would intersect {\it only once} with each layer of the tracker. Therefore, particles satisfying criteria for stub formation ($p_t>2 \ \rm GeV/c$) would also intersect only once with each layer. Let us consider a particle produced at collision point with momentum $p_x=p_t, \ \rm and \ p_y=p_z=0$, thus, traveling along the x-axis with an initial azimuth angle of $\phi_0=0$. Center of the circular trajectory for initial momentum $(p_t,0,0)$ is $(0, R_p)$, where radius of curvature $R_p$ is obtained using Eqn. \ref{eqn:radius}. It can be shown that for a particle traveling along x-axis at production ($\phi_0=0$),

\begin{equation}\label{eqn:Rc}
 R_p =  \frac{(R_c^L)^2 } {2 \times y_c^L} = \frac{R_c^L}{2 \times Sin \delta_c^L} \hskip 0.4cm {\rm for} \ \phi_0=0
\end{equation}

where, a) $R_c^L$ is radius of the cylindrical layer {\it L} (see Fig. \ref{fig:tracker}), b) $y_c^L$ is the y-coordinate of the intersection point of trajectory with the same layer and c) $\delta_c^L$ is azimuthal angle which is actually a deflection angle of the trajectory since $\phi_0=0$. In above equation $y_c^L$ is the only dynamic parameter using which angle of deflection and radius of curvature can be calculated. Using Eqn. \ref{eqn:radius} and \ref{eqn:Rc} it can be shown that,

\begin{equation}\label{eqn:delta}
 Sin\delta_c^L =  \frac{\beta \times R_c^L} {p_t} \ \ {\rm where, } \hskip 0.7 cm \beta = 0.15 \times B_z
\end{equation}

\noindent Using this Equation, the difference of Sine of deflection angle between successive layers can be obtained as,

\begin{equation}\label{eqn:diff}
 \Delta S\delta_c^{L+1,L} = Sin\delta_c^{L+1} - Sin\delta_c^{L} = \frac{\beta \times (R_c^{L+1} - R_c^L)} {p_t} 
\end{equation}

\noindent With this, we introduce the filter parameter $\alpha$ as,

\begin{equation}\label{eqn:alpha}
 \alpha_{L+2,L} = \frac {\Delta S\delta_c^{L+2,L+1}}{\Delta S\delta_c^{L+1,L}} = \frac{R_c^{L+2} - R_c^{L+1}} {R_c^{L+1} - R_c^{L}} 
\end{equation}

It can be seen that the middle term in Eqn. \ref{eqn:alpha} is estimated using dynamic parameters of the potential tracklet (triplet) candidates, {\it i.e.,} using deflection angle ($\delta_C^L$) of stubs in consecutive three layers, whereas, the last term is evaluated using the static parameters of the detector geometry, {\it i.e.,} radii of the corresponding cylinders, which is independent of the triplet parameters. Henceforth, the filter parameters calculated using tracklet parameters ($\delta_C^L$) would be referred as  $\alpha_{L+2,L}^{obs}$, whereas, the same calculated using the radii of the corresponding cylinders ($R_c^L$) would be  referred as  $\alpha_{L+2,L}^{exp}$. The filter parameter of the triplet candidate, $\alpha_{L+2,L}^{obs}$, is compared with corresponding $\alpha_{L+2,L}^{exp}$ to qualify triplets. Accepted triplets are progressively propagated to subsequent layers to build complete tracklet. The tracklet  if build successfully, is expected to corresponds to the stubs of one of charged the particle trajectory produced in collision.

\subsection{Track Reconstruction using Filter Parameters at $\boldsymbol{\phi_0=0}$}\label{sec:fp1}

In this section, we describe the effective use of  filter parameters ($\alpha_{ij}$) introduced in Eqn.\ref{eqn:alpha} for reconstructing tracks that generated with varying $p_t$ and fixed azimuth angle $\phi_0=0$. A consecutive three layer tracklets ($T_{i,i+1,i+2}$) are formed using the corresponding filter parameters. Exact procedure for mapping stubs to tracklets and subsequently to their respective trajectories for positively charged  particles ($\delta_c^{L+1}>\delta_c^{L}$) with $\phi_0=0$ ($p_y=0$) is described below. 

\begin{enumerate}
    \item Consider a stub from first layer having deflection angle $\delta_c^1$. 
    
    \item In Layer $2$, select only those stubs with deflection angle $\delta_c^2$ satisfying $0^0<\delta_c^2-\delta_c^1<5^0$. It is to be noted that deflection angle for lowest $p_t^{min}=2 \ \rm GeV/c$ track is $<5^0$.
    
    \item Select stubs in Layer $3$ with deflection angle $\delta_c^3$ satisfying $0^0<\delta_c^3-\delta_c^2<5^0$. Using deflection angles $\delta_c^1, \ \delta_c^2 \ \rm and \ \delta_c^3$, calculate observed $\alpha_{31}^{obs}$ as shown in Eqn.~\ref{eqn:alpha}. Compare the observed $\alpha_{31}^{obs}$ with the expected $\alpha_{31}^{exp}$ calculated using radii of cylindrical layers $1$, $2$ and $3$. Stubs are accepted to form a tracklet ($\rm T_{123}$) if and only if $\alpha_{31}^{obs} \ \rm and \ \alpha_{31}^{exp}$ are consistent with each other, i.e., $\Delta\alpha_{31}=|\alpha_{31}^{obs}-\alpha_{31}^{exp}|<0.01$. This is quite powerful constrain and rejects almost all the fake combinatorics and retains only those stubs, corresponding to its original track. 

    \item Using stubs in the tracklet $\rm T_{123}$, obtain the successive tracklet  $\rm T_{234}$ by finding a stub in layer $4$ such that,  $0^0<\delta_c^4-\delta_c^3<5^0$ followed by  $\Delta\alpha_{42} <0.01$ criteria. This is done by using the same procedure as described in Step $3$. It is to be noted that, stubs from Layer $2$ and $3$ in $\rm T_{234}$ are same as that in $\rm T_{123}$
    
    \item Using same approach as described above, successive tracklets $\rm T_{345}$ and $\rm T_{456}$ are obtained.
    
\end{enumerate}

\begin{figure}[!h]
    \centering
    \includegraphics[width=0.95\linewidth]{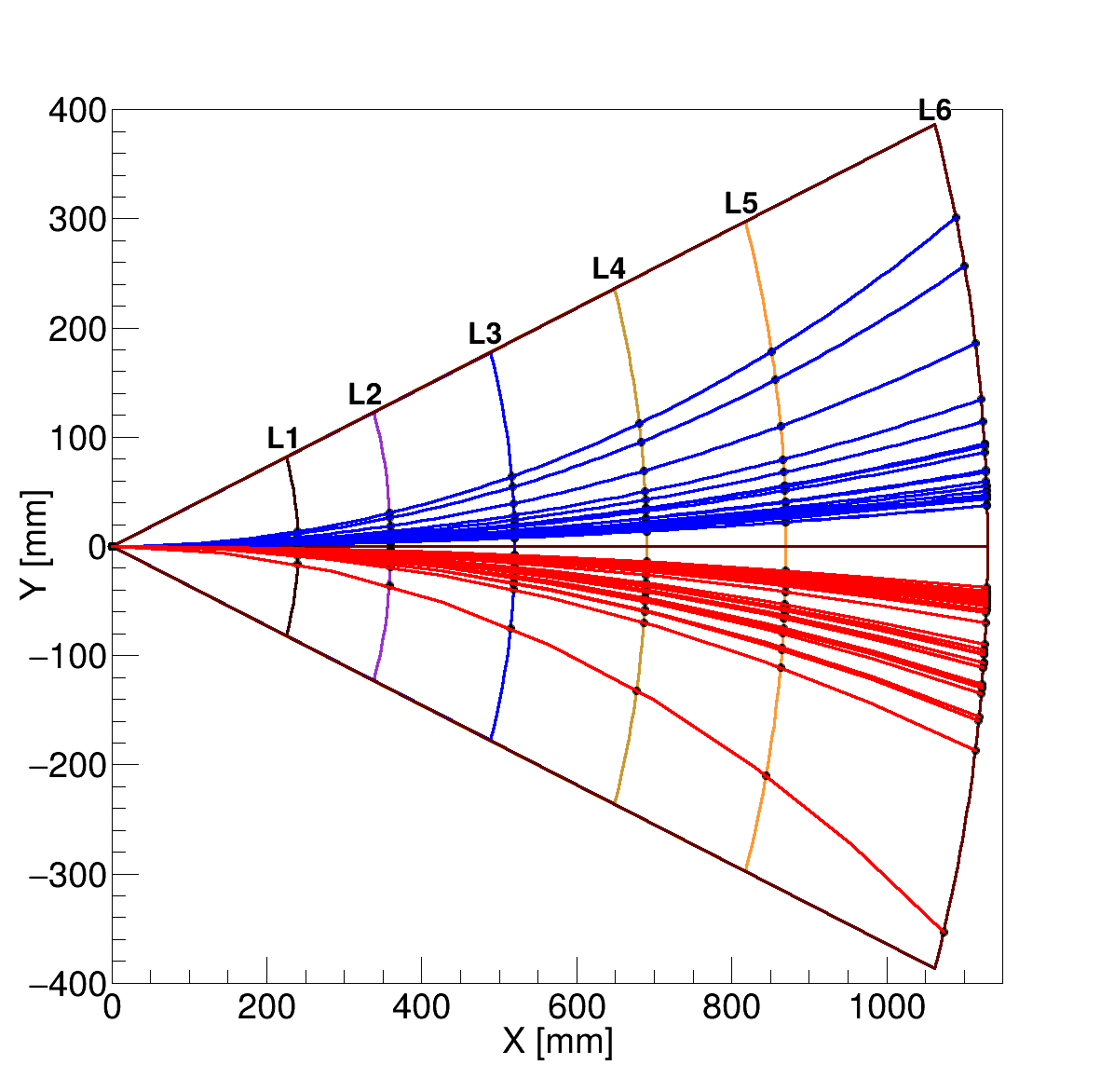}
    \caption{Reconstructed tracks along with stubs for {\it geometrically} propagated charged tracks ($2<p_x<20 \ GeV/c, \ q=\pm1$)  at $\phi_0=0$ projected in the transverse plane of the outer tracker (Blue: q=+1 and Red: q=-1)}
    \label{fig:reco_fixed}
\end{figure}

Stubs corresponding to the particle trajectory are obtained by combining stubs in all the four tracklets that are build sequentially by applying stringent criteria provided by the filter parameters. Subsequently, track's kinematic parameters are obtained by using stub coordinates. For negatively charged particles, the difference in deflection angle condition is reversed as $0^0<\delta_c^L -\delta_c^{L+1}<5^0$. Fig. \ref{fig:reco_fixed} shows the stubs and reconstructed tracks obtained using filter parameters for fifty particles ($q=\pm 1$) that are propagated {\it geometrically} through the tracker with varying $p_x$ ($2<p_x<20 \ GeV/c, \ p_y=0 \ \Rightarrow \phi_0=0$). It can be seen that proposed algorithm based on filter parameter builds all the true tracks with no contribution due to the fake combinatorics.

\subsection{General formalism for Track Reconstruction using Filter Parameters}\label{sec:fp2}
Though the filter parameter based mechanism performs exceedingly well; however, the formalism is strictly applicable to only those tracks with $p_y=0$, i.e. $\phi_0=0$ as described in previous Section \ref{sec:fp1}. Same formalism cannot be used  for tracks with $\phi_0>0$ because the value of the filter parameter depends on the azimuth angle $\phi_0$ at production and therefore it looses its static characteristic identified {\it uniquely} with the radii of the tracker cylinders (Eqn. \ref{eqn:alpha}). Dependence of $\alpha_{ij}$ on $\phi_0$ has been studied in detail by generating large number of tracks at fixed $p_t$ ($p_z=0$)  with $\phi_0$ varying between $0^0 \le \phi_0 \le 360^0$. These tracks are propagated {\it geometrically} through the tracker geometry shown in Fig. \ref{fig:tracker}. The filter parameters are calculated for each track using their azimuthal angles ($\phi_i, \ i=1 \ \rm to \ 6$) for each layer. Various features of dependence of $\alpha_{31}$ on $\phi_0$ observed from the generated data are listed below. 

\begin{figure*}[!h]
    \centering
    \includegraphics[trim=2pt 2pt 2pt 2pt,clip,width=0.48\linewidth]{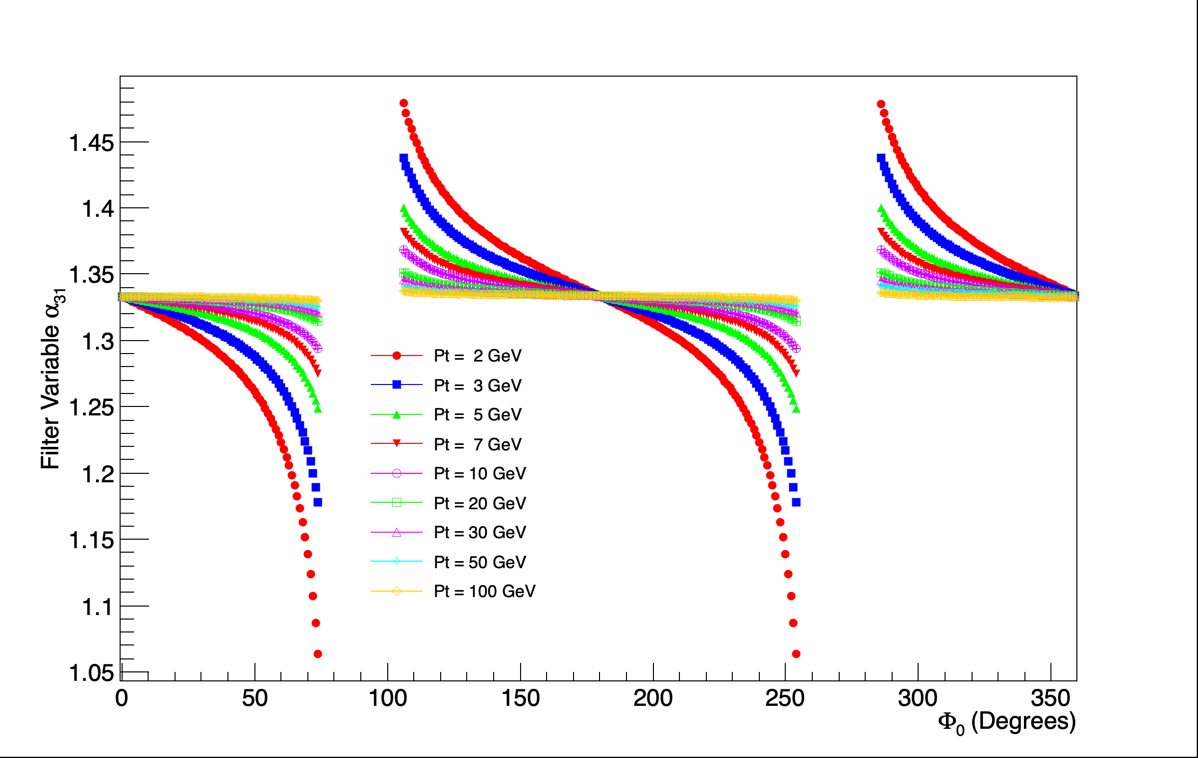}
    \includegraphics[trim=2pt 2pt 2pt 2pt,clip,width=0.48\linewidth]{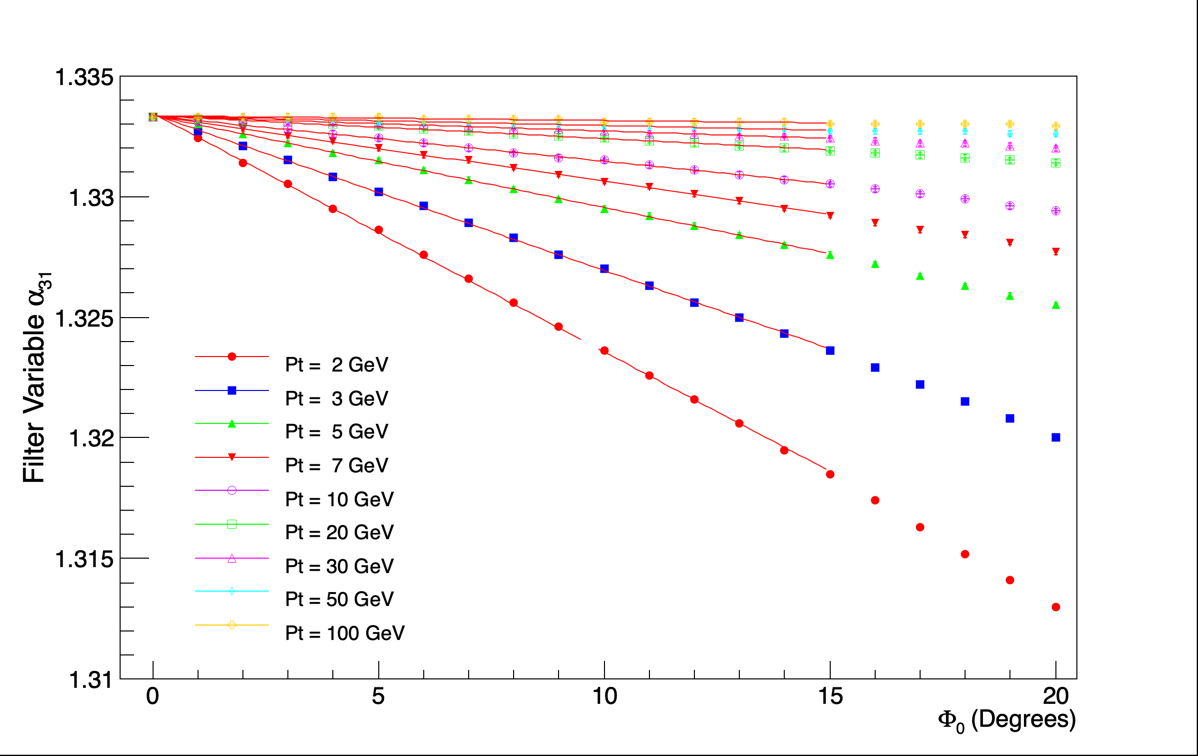}
    \caption{Variation of filter parameter $\alpha_{31}$ {\it w.r.t.} azimuth angle in $0-360^0$ (Left) 
    and $0-20^0$ (Right) range}
    \label{fig:alpha}
\end{figure*}

\begin{enumerate}

  \item Figure \ref{fig:alpha} (Left) shows variation of $\alpha_{31}$ as a function of $\phi_0$ for different values of $p_t$. Variation in $\alpha_{31}$ also depends on the $p_t$ of the track. 

  \item The value of $\alpha_{31}$ is independent of $p_t$ for tracks having $\phi_0=0$
  
  \item The $\alpha_{31}$ varies rapidly with $\phi_0$ for lower $p_t$ tracks and moderately for higher $p_t$ tracks.

  \item The $\alpha_{31}$  diverges at $\phi_0=90^0 \ \rm and \ 270^0$, hence a region of $15^0$ around the singularities is not shown in the figure for a better clarity.

  \item Similar behavior is seen for other filter parameters such as,  $\alpha_{42}, \ \alpha_{53} \ \rm and \   \alpha_{64}$.
\end{enumerate}

  Due to large variation in $\alpha_{ij}$ as a function of $\phi_0$, including divergences, it can not be used as it was successfully exploited for mapping stubs to their respective trajectories with $\phi_0=0$. However, as can be seen from Figure \ref{fig:alpha} (Right), the $\alpha_{31}$ varies within a very small range (0.5\%) in the low $\phi_0$ region ($0^0<\phi_0<5^0$) for any $p_t$. Furthermore, maximum deflection angle of tracks between successive layers is $<5^0$ for $p_t >2 \ GeV/c$.  Hence, one of the possible way to get back to the ideal case, is to use the rotational symmetry of the tracks in the transverse plane. Procedure for using filter parameter based mechanism for $0 \leq \phi_0 \leq 360$ case is explained below:

\begin{enumerate}

\item Find an offset angle $\phi_{offset}$ such that $\phi_1^{'} = \phi_1-\phi_{offset}=5^0$. This ensures that $\phi_0^{'}=\phi_0-\phi_{offset}$ will be close to $0^0$, hence, enabling us to use the same formalism described in Section \ref{sec:fp1}.

\item Apply the offset angle $\phi_{offset}$ on azimuth angle of all stubs of other layers, $i.e., \ \phi_L^{'} = \phi_L-\phi_{offset} \ \ \rm for \ L=2 \ to \ 6$

\item For a given stub in layer 1, form three layer tracklet $T_{123}$ by identifing stubs in layer 2 and layer 3 satisfying $0< \phi_2' - \phi_1' < 5$ and $0< \phi_3' - \phi_2' < 5$.

\item Estimate $\alpha^{obs}_{13}$ by substituting $\phi_1^{'}, \ \phi_2^{'}, \ \phi_3^{'}$ ($\phi_i' \equiv \delta_c^i$) in Eqn. \ref{eqn:alpha}. If stubs belong to the same particle's trajectory then the observed $\alpha^{obs}_{13}$ would be very close to the expected $\alpha_{13}^{exp}$ estimated using the radii of the corresponding cylindrical layers. 

\item Accepted tracklets $T_{123}$ are used to find subsequent tracklets ($T_{234}$, $T_{345}$ and $T_{456}$) that are validated by corresponding filter parameters. Using them, a complete tracklet ($T_{123456}$) is build representing stubs one of the particle's trajectory.

\end{enumerate}

\begin{figure}[!h]
    \centering
    \includegraphics[width=0.95\linewidth]{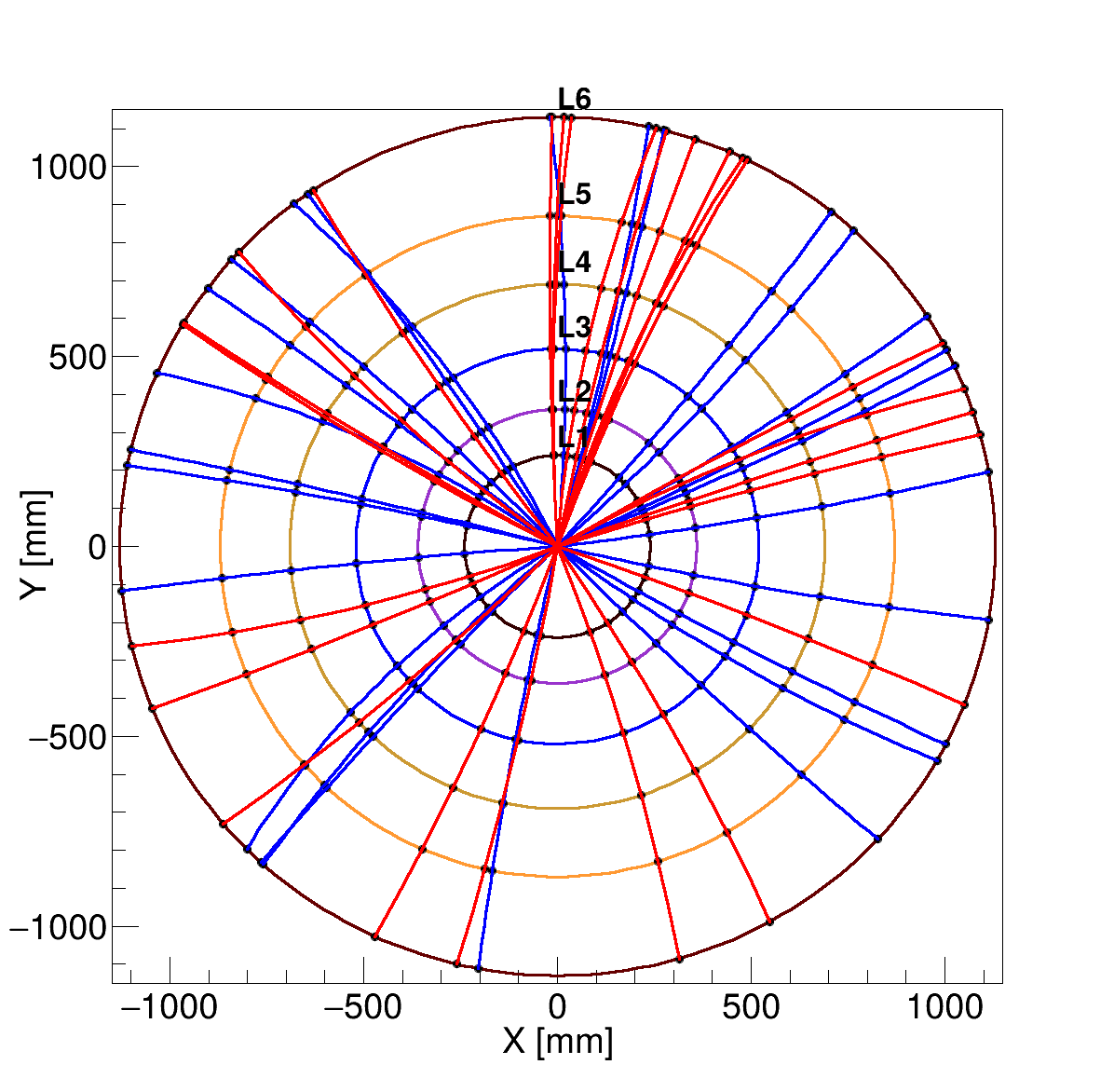}
    \caption{Reconstructed tracks along with stubs ($\rm q=\pm1, \ \ 2 <p_t<20 \ GeV/c \ \ and \ \ 0<\phi_0<360^0$) projected in transverse plane of the tracker}
    \label{fig:reco_any}
\end{figure}

Fig. \ref{fig:reco_any} shows stubs and reconstructed tracklets for fifty tracks with $\rm q=\pm1, \ 2 <p_t<20 \ GeV/c \ \ and \ \ 0< \phi_0<360^0$ that are obtained using the filter parameter based mechanism. 

So far, a full tracklet formation pipeline based on filter parameter approach has been explained and demonstrated by using few ($\sim 100$) particles that are {\it geometrically} projected into the tracker layers with $\eta_0=0$. However, during HL-LHC, large number of particles ($\sim2K$) with varying $p_t, \eta_0 \ \rm and \ \phi_0$ are produced in each bunch crossing event containing large number of pileup collisions. Following section describes the simulation of pp collisions for top quark pair ($t\bar t$) production along with a large number of pileups (soft QCD) forming a bunching crossing event for HL-LHC era at $\sqrt s= 14 \ TeV$. It is followed by reconstruction of tracklets (Section \ref{sec:reco_pythia8}) of geometrically projected trajectories of charged particles produced in each bunch crossing event.

\section{Event Generation Framework}\label{sec:evt_gen}

Monte Carlo (MC) event generators are essential tools in any collider physics experiments used for simulating relativistic particle collisions and modeling detector responses. Widely used generators such as \textsc{Pythia8} ~\cite{Sjostrand:2007gs} and \textsc{Herwig}++ ~\cite{Bahr:2008pv} provide detailed descriptions of collision final states, including parton showering, hadronization, and multi-parton interactions. In this study, \textsc{Pythia8} was used to generate large number of pileup and $t\bar{t}$ pair production events separately for $pp$ collision at $\sqrt{s}=14~\mathrm{TeV}$. Jets were reconstructed using the anti-$k_t$ clustering algorithm implemented in the \textsc{FastJet} package ~\cite{Cacciari:2011ma}. Generated particles and reconstructed jets, along with their kinematic and production parameters were stored using the \textsc{ROOT} framework ~\cite{Brun:1997pa}. 10K events each of pileup and $t\bar{t}$ events are generated using \textsc{Pythia8}. Using this dataset, bunch crossing events corresponding to HL-LHC conditions are generated. Following section describes detailed procedure for generating bunch crossing events.

\subsection{Simulation of Bunch Crossing Events at HL-LHC}\label{sec:sim_bunch}
At the LHC, each bunch crossing triggered event typically corresponds to a large number of pileup (PU) interactions along with a hard interaction process. The number of PU interactions in each bunch crossing depends on the instantaneous luminosity, and it can go up to as high as 200 interactions during the HL-LHC era. Therefore, simulation of bunch crossing event essentially corresponds to combining several pileups and a hard process interaction event ($t\bar t$ production in our case). Bunch crossing event formation steps are described below:

\begin{enumerate}
    \item For a given bunch crossing event, generate number of pileup events, $N_{PU}$, using a gaussian distribution with mean $<nPU>=150$.

    \item Randomly pick $N_{PU}$ events from the pileup dataset and a single $t\bar t$ event from the hard interaction dataset. 
    
    \item Club $N_{PU}$ pileup events and a single $t\bar t$ event together to form a bunch crossing event. Composed event, consists of large number of particles along with their complete kinematic information.

    \item Repeat above process for subsequent hard interaction events. A total of 1K bunch crossing events are generated using this approach and stored in the \textsc{ROOT} based dataset. This dataset is subsequently used for \textsc{Geant4} simulation (Section \ref{sec:geant4}).

\end{enumerate}

The pileup collision being dominated by soft QCD processes have a smaller mean charge multiplicity and a steeper fall-off in the $p_t$ distribution as compared to hard interaction ($t\bar t$ production). However, during the HL-LHC era, since each bunch-crossing will be having a large number of pileup interactions, the recorded events are expected to be dominated by pileup tracks. 

\subsection{Track Reconstruction for HL-LHC Phase using Generator Level Data} \label{sec:reco_pythia8}

In this section, the reconstruction of generator level particles, {\it geometrically} propagated through the outer tracker toy detector geometry is demonstrated using the proposed filter parameter based approach.  Each bunch crossing event typically consists of about 2K charged particles with $p_t>2 \ \rm GeV/c$ and $|\eta|<3$. Each charged particle using its kinematic parameters is propagated  {\it geometrically} from its vertex through the outer tracker system of six layers in the presence of a longitudinal magnetic field of 3.8 Tesla, resulting in about 12K stubs, provided each particle goes through all six layers of the tracker. Azimuthal and longitudinal pixel indices ($\phi_i^L, \ z_i^L$) of all stubs from each layer are mapped to their corresponding tracks using filter parameters based tracklet reconstruction algorithm. As per one of the the design parameters of the accelerator, vertex of each pileup and hard interaction in bunch crossing event is fluctuated around z=0 using a gaussian distribution with $\sigma_z=4 \ cm$ ~\cite{Collaboration_2014}. Impact of this is linearly translated to the longitudinal pixel index ($z_i^L$) of all stubs. Vertices are fluctuated in the range of $\pm3.5\sigma_z$. Complete procedure for formation of full tracklets, i.e., identifying stubs associated with each of the tracks, is described in detail below:

\begin{enumerate}

\item Using a stub each from two consecutive innermost layers (Layer 1 and 2), a straight line trajectory criterion in the r-z plane is applied with a requirement of its intersection with the beam axis within $\pm3.5\sigma_z$. It helps in reducing fake combinatorics while identifying all possible doublets ($T_{12}$)

\item The deflection angle ($\Delta \phi_{L,L+1}$) of a trajectory between successive layers in $r-\phi$ plane is inversely proportional to the transverse momentum, $p_t$, of the particle. Hence, for a particle with a minimum $p_t$ of 2 GeV/c, the maximum deflection angle between successive layers are: $\Delta \phi_{12}=1.97^0$, $\Delta \phi_{23}=2.63^0$, $\Delta \phi_{34}=2.82^0$, $\Delta \phi_{45}=3.02^0$ and $\Delta \phi_{56}=4.43^0$. The tracklets ($T_{12}$) passing through the selection criteria in Step 1 are further subjected to the maximum deflection angle criteria in the $r-\phi$ plane with an additional tolerance of 30\% on maximum deflection angles. Also, consistency in the direction of deflection between successive layers (depending on the charge state) is ensured while applying this condition. The deflection angle criteria further eliminates the fake combinatorics in a significant manner.

\item Two layer tracklets ($T_{12}$) formulated using Steps 1 and 2 are extended to form three layer tracklets ($T_{123}$) with stubs from Layer 3. Three layer tracklets are subjected to a similar selection criteria described in Steps 1 (straight line in r-z plane) and 2 ($\Delta \phi_{23} < 1.3 \times 2.9^0$). These tracklets are further subjected to the filter parameter criteria for $\alpha_{31}$ as explained in Sections \ref{sec:fp1} and \ref{sec:fp2}. It is to be noted that the number of $T_{123}$ obtained after applying all the three selection criteria are much smaller than that of $T_{12}$ due to higher elimination of fake combinatorics in $T_{123}$ with almost $100\%$ retention of true trajectories.

\item Four layer tracklets ($T_{1234}$) are  obtained by projecting accepted three layer tracklet $T_{123}$ into the Layer 4. These tracklets ($T_{1234}$) are subjected similar conditions as described in Steps 1 and 2 ($\Delta \phi_{34} < 1.3 \times 2.82^0$) and filter parameter criteria for $\alpha_{24}$.

\item Procedure described above for the formation of tracklets $T_{1234}$ is repeated to form five layer tracklets ($T_{12345}$) followed by six layer tracklets ($T_{123456}$).

\end{enumerate}

\begin{figure*}[h!]
\centering
\includegraphics[width=0.45\linewidth]{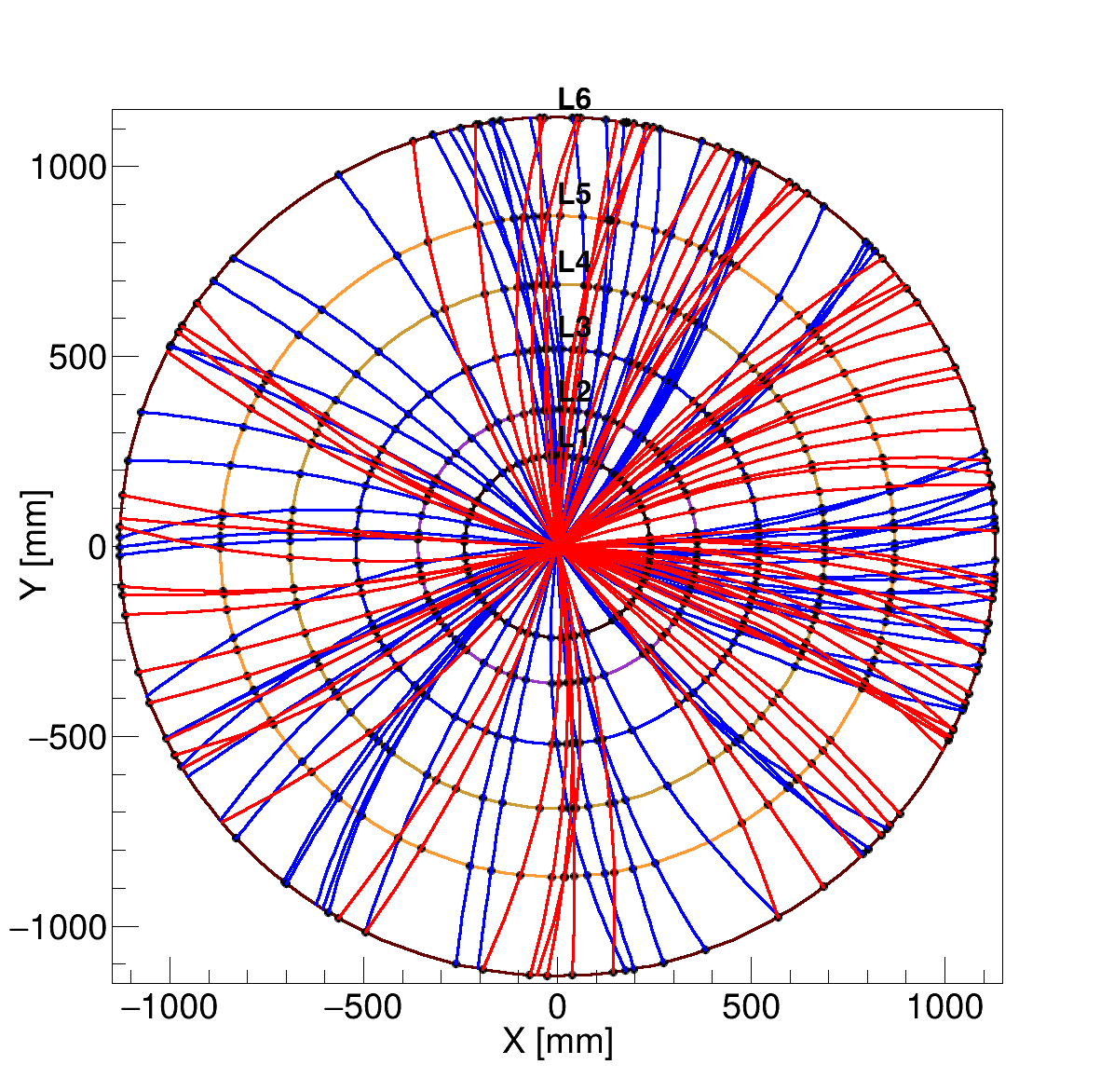}
\includegraphics[width=0.45\linewidth]{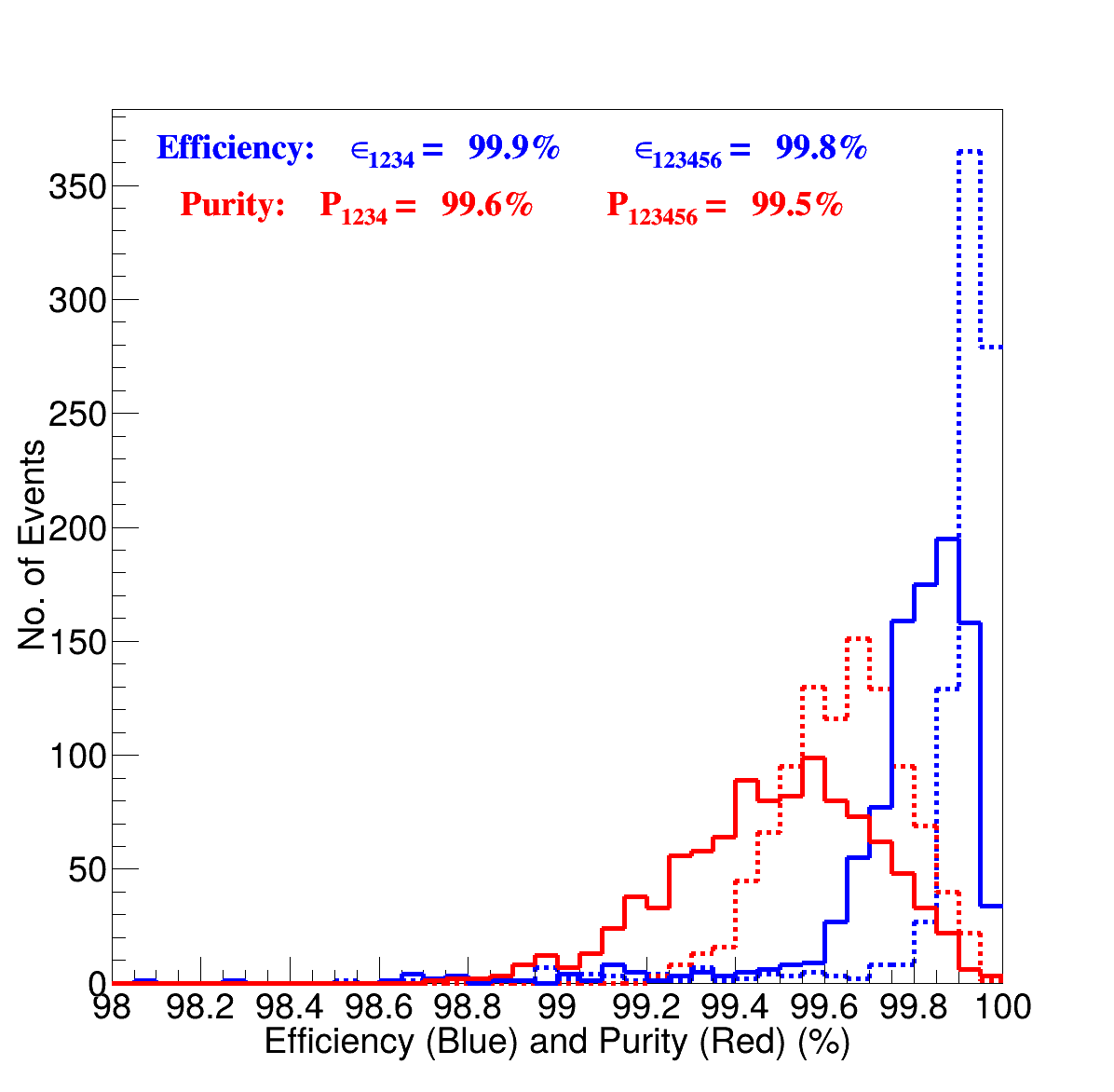}
    \caption{Left:  Tracks projected in transverse plane reconstructed using mapped geometrically projected hits (Blue: q=+1 and Red: q=-1). Right: Efficiency (blue) and purity (red) of reconstructed tracks for 4 (dotted) and 6 (solid) layer tracklets.}
    \label{fig:pythia8_tracks}
\end{figure*}

This completes the tracklet formation process mainly based on three constraints a) straight line criteria in r-z plane and intersection with beam axis within $\pm 3.5\sigma_z$, b) upper limit on deflection angle between successive layers and c) consistency between observed and expected filter parameters ($\alpha_{ij}^{obs}$ and $\alpha_{ij}^{exp}$). Same procedure is used for both the charge states ($q=\pm1$). Figure \ref{fig:pythia8_tracks} (Left) shows the reconstructed tracks in transverse plane for 150 particles out of about 2K reconstructed tracks from one of the bunch crossing events. A smaller number of tracks are drawn to have better visibility of the reconstructed bunch crossing event. As can be seen from the Figure \ref{fig:pythia8_tracks} (Left), the mapping between stubs and their associated track is excellent. There are no visible, poorly reconstructed tracks in the figure. In order to quantitatively understand the efficacy of the proposed algorithm, we study the purity (P) and efficiency ($\epsilon$) of tracklet formation for each type of tracklet. For a given bunch crossing event, these parameters (P and $\epsilon$) for a three-layer tracklets ($T_{123}$) are defined as: 

\begin{equation}\label{eqn:reco_eff}
\begin{aligned}
 \epsilon_{123} &= \frac {N_{123}^{miss}}{N_{123}^{total}} \\ 
 P_{123} &= 1- \frac {N_{123}^{fake}}{N_{123}^{total}} 
\end{aligned}
\end{equation} 

where $N_{123}^{total}$  are the total number of tracks passing through at least three layers of tracker, $N_{123}^{miss}$ and $N_{123}^{fake}$ are the number of missed (not reconstructed) and fake (wrongly reconstructed) tracklets, respectively, in a given bunch crossing event. The efficiency and purity are obtained for each type of tracklets, ranging from $T_{123}$ to $T_{123456}$, and their distributions for tracklets up to four and six layers are shown in the Figure \ref{fig:pythia8_tracks} (Right). As can be seen, the proposed algorithm exhibits outstanding performance by providing very high efficiency and purity ($>99 \%$) thereby confirming its potential for its application in the HL-LHC era.

However, so far this algorithm has been demonstrated only on geometrically propagated tracks through the tracker. Passage of charged particles through passive and active material of the tracker results into energy loss and also production of several secondary particles that can change the performance of the proposed track reconstruction algorithm. Thus, it is important to simulate realistic collisions at the HL-LHC that are typically recorded in the experiment. 
In view of this, a more realistic response of the tracker, in presence of high magnetic field, to a large number of charged particles is simulated using \textsc{Geant4} based ~\cite{GEANT4:2002zbu} framework described in the following section ~\ref{sec:geant4}.

\section{Simulation of Tracker using \textsc{GEANT4} Framework }\label{sec:geant4}
The \textsc{Geant4} framework simulates the passage of particles through a prototype outer tracker geometry immersed in a  longitudinal magnetic field of 3.8 Tesla. Key design parameters of the barrel outer tracker are kept similar to the CMS tracker ~\cite{CERN-LHCC-2017-009} to be operated during the HL-LHC era. First step towards reconstruction of individual tracks, under high pileup environment, is to accurately identify stubs associated  with each of the tracks (tracklet formation) using the proposed algorithm, with high efficiency and purity. Stubs associated with each tracklet are fed to the machine learning (ML) model to precisely estimate the track's transverse momentum vector, and its charge (Section \ref{sec:MLT}). Construction of the prototype outer barrel tracker geometry and simulation of passage of particles through the tracker using the \textsc{Geant4} \cite{GEANT4:2002zbu} framework is described in this section. The \textsc{Geant4} framework has advanced capabilities in implementing detector geometry and material properties. The framework accurately models particle's trajectory and interactions with matter.

The prototype tracker geometry that is similar to the CMS outer tracker geometry is constructed using the geometry modules of \textsc{Geant4} framework. The tracker consists of six cylindrical concentric layers (Fig \ref{fig:tracker}) of active silicon of thickness $300 \ \mu m$ appropriately segmented to record the hits generated by passage of charged particles through these layers. Hit coordinates are digitized with a spatial resolution of pixels matching with the CMS outer tracker design. Energy loss of hits mapping to the same pixel are summed up to get the energy loss in a pixel. Schematic of the outer tracker and detector dimensions are shown in Figure \ref{fig:tracker}. The kinematic parameters  of each particle in a bunch crossing event are used to simulate their trajectory through the detector geometry using tracking modules of the \textsc{Geant4} framework. Key parameters of each hit, such as a) particle ID (pID), b) its energy before entering ($E_b$) and after exiting ($E_a$), and c) its energy loss ($\Delta E$) in each active layers along with its longitudinal and azimuthal position ($\rm pID, \ L, \ z_L, \ \phi_L, \ E_b, \ E_a, \ \Delta E$) are stored. The energy loss in a pixel is obtained by summing over $\Delta E$ of each hit whose position maps to the same pixel. This is the most realistic dataset, representing the HL-LHC era, which has been used for complete reconstruction. One thousand bunch crossing events (each consisting of about 150 pileups and a single hard interaction) generated using \textsc{Pythia8} are simulated through the tracker detector geometry using the \textsc{Geant4}framework described above.

\subsection{Validation of Tracker Geometry} \label{sec:geant4-valid}
It is important to validate the response of the detector geometry in the presence of a magnetic field before deploying it to simulate large number of bunch crossing events. To qualify the detector response and the geometry, a large number of single muons ($q=\pm1$) with fixed transverse momentum ($p_t=3 \ \rm GeV/c$) and  varying pseudorapidity ($\eta_0$) and azimuth angle ($\phi_0$) are simulated from the vertex through the tracker geometry in presence of a longitudinal magnetic field of $B_z = 3.8 \ \rm Tesla$.  Muons are chosen to validate the geometry as they primarily undergo energy loss due to the ionization process, thus avoiding the complexity of showering effects of energy loss due to bremsstrahlung or hadronic processes enabling a reliable benchmark in assessing the detector response. 

\begin{figure}[!h]
\centering
\includegraphics[width=0.99\linewidth]{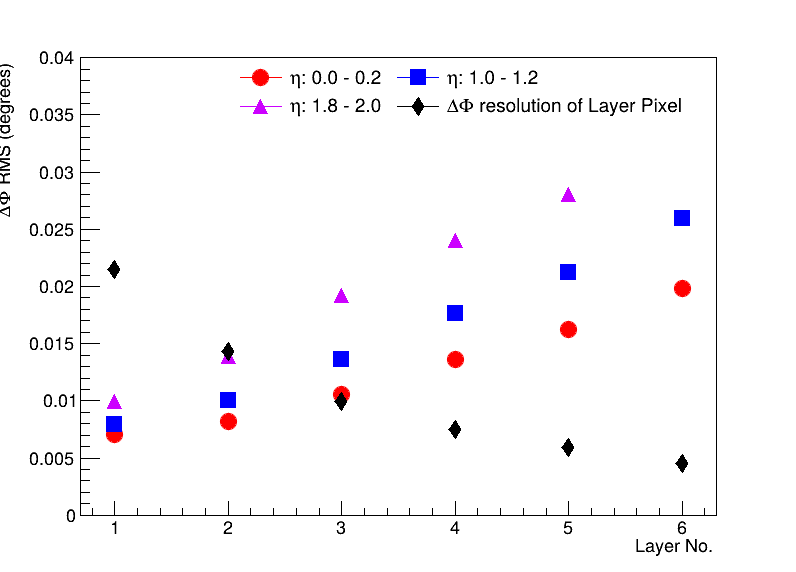}    
\caption{Scattering angle in transverse plane as a function of Layer number in different $\eta$-regions. Resolution of each layer pixel in the azimuthal direction is also shown (black diamond).} 
\label{fig:geant4-muon-valid}
\end{figure}

Simulated hits are digitized as described in Section \ref{sec:geant4}. Pixels with incoming $p_t<2 \ \rm GeV/c$ do not satisfy the stub criteria and hence rejected. Simulated distributions of energy loss by muons is seen to follow a) the Landau distribution with its peak agreeing to the expected value and b) the energy loss in pixels increases with $\eta_0$ of particle at vertex due to increase in the path length of muon through the active layers, thus, validating the energy loss response of the active component (silicon layers) of the detector. Distribution of angular difference between simulated and expected position of the track in each layer is consistent with expected scattering angle with a mean of zero, validating the geometry of the detector. The width of the distribution represents the scattering angle due to multiple coulomb scattering. It depends on the length and properties of the material that the track has passed through before entering the pixel. Hence, the scattering angle is expected to increase with layer number and pseudorapidity. The scattering angle plotted as a function of layer number in different $\eta$ ranges is shown in Figure \ref{fig:geant4-muon-valid}. It follows the expected trend, thereby validating the simulation of the energy loss mechanism. Match between expected and observed pixel positions validates the detector geometry. 
It is to be noted that the inner two layers have a coarser azimuthal pixel resolution as compared to the scattering angle of these layers, whereas, outer four layers has better azimuthal pixel resolution as compared to the scattering angle of these layers.

\subsection{Uncertainties on Filter Parameters}\label{sec:geant4-filter} 

As described in the previous section, projected and simulated trajectories do not necessarily follow the same path. Projected ({\it true}) hit positions of particles are obtained by the geometrical projection of the track using its kinematic parameters and the magnetic field. However, simulated hit positions have a spread around the {\it true} hits due to finite pixel size, energy loss and coulomb scattering resulting in a small deviation of the simulated trajectory {\it w.r.t.} the {\it true} trajectory. Figure \ref {fig:geant4-muon-valid} shows the contribution to the angular resolution from coulomb scattering as well as the layer-wise pixel resolution.

\begin{figure*}[!h]
\centering
\includegraphics[width=0.48\linewidth]{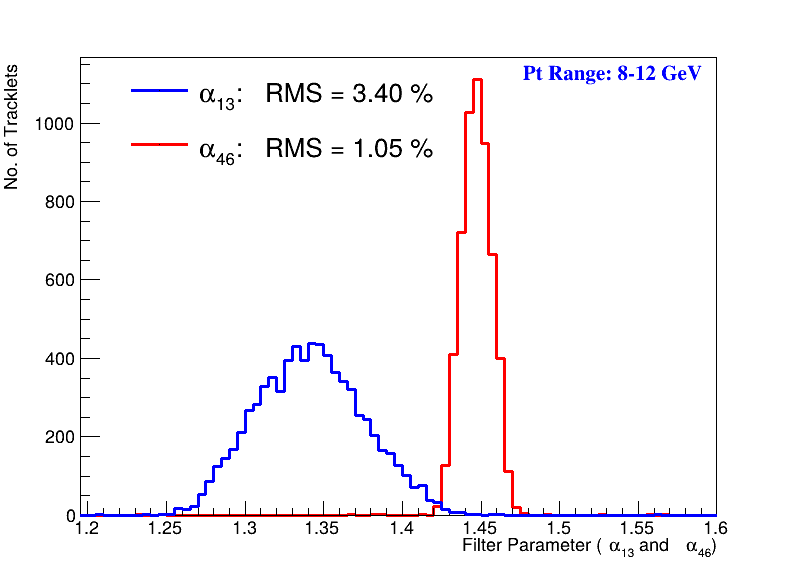}
\includegraphics[width=0.48\linewidth]{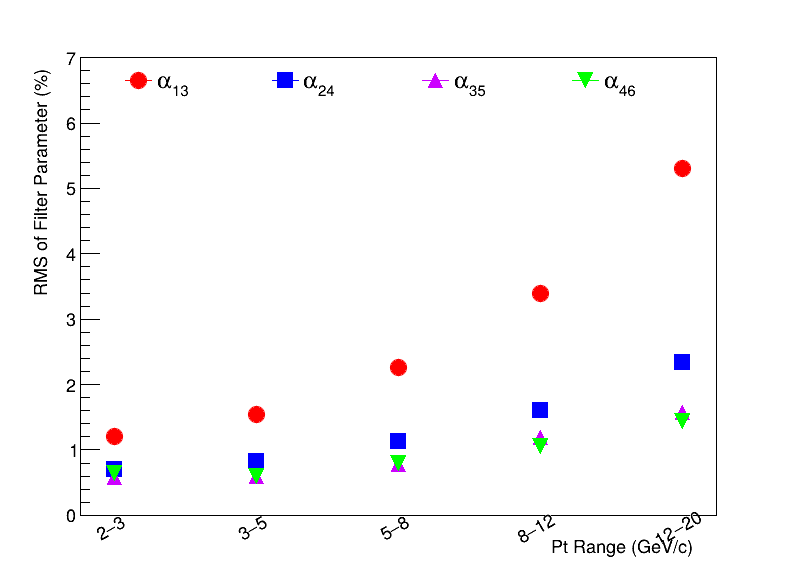}
\caption{Left: Distribution of filter parameters ($\alpha_{13}, \ \alpha_{46}$) for simulated tracks for a $p_t$ range of $8 < p_t < 12 \ \rm GeV/c$. Right: RMS (\%) of distinct filter parameters as a function of $p_t$ ranges.} 
\label{fig:geant4-hadron-valid}
\end{figure*}

The filter parameters (FP) proposed in Section \ref{sec:fp}, $\alpha_{ij}^{obs}$, are estimated using the azimuthal angle of simulated pixel hits from any consecutive three layers using Eqn. \ref{eqn:alpha}. Inaccuracies in the filter parameters estimated using these hits depend on the respective uncertainties in their recorded azimuthal angles. Hence, it is important to estimate the inaccuracy of FP, which can then be used while filtering uncorrelated hits during tracklet formation process to optimize reconstruction efficiency  with minimal fake contributions. The FP estimated for each of the simulated particles is obtained using their associated simulated pixel hits azimuthal position. The distributions of FP $\alpha_{13}^{obs}$  (inner three layers) and $\alpha_{46}^{obs}$ (outer three layers) in a certain $p_t$ range ($8 < p_t < 12 \ \rm GeV/c$) are shown in Figure \ref{fig:geant4-hadron-valid} (Left). Due to a better intrinsic azimuthal angular resolution of outer layer pixels as compared to the inner layers, the FP $\alpha_{46}^{obs}$ has a sharper distribution as compared to $\alpha_{13}^{obs}$ in the same $p_t$ range. Using similar distributions, the RMS for each filter parameter are obtained in different $p_t$ ranges of simulated particles, shown in Figure \ref{fig:geant4-hadron-valid} (Right). It is to be noted that the mean of a given filter parameter is observed to be independent of the $p_t$ of particle. However, the RMS of FP, i.e., inaccuracies in the FP increases with increasing $p_t$ since trajectories with high $p_t$  have a smaller deflections as compared to the azimuthal resolution of the pixels. Tracklets formed using consecutive three layers are rejected if their corresponding FP estimated using azimuth angle of pixel hits is outside  the expected acceptance window ($\pm \ 3 \times RMS(\alpha_{ij}^{exp})$). 

\subsection{Tracklet Formation for \textsc{Geant4} Based Simulation Data at HL-LHC}\label{sec:reco_geant4}

In Section \ref{sec:reco_pythia8} association of projected stubs with their corresponding expected trajectories in the tracker was demonstrated. However, the efficacy of the filter parameter has to be demonstrated on the simulated data that closely replicates the real data. It is to be noted that projected stubs obtained by {\it geometrically} projected tracks from the primary vertex to the outer tracker do not take into account the energy loss of particles in the active and passive medium of the detector. In light of this, passage of all the particles produced in each bunch crossing are simulated through the outer tracker geometry using the \textsc{Geant4} framework (Section \ref{sec:geant4}).  These hits are digitized and only those pixel hits with incoming particles with $p_t > 2 \ \rm GeV/c$, as identified by the L1 Trigger, are considered as stubs for tracklet formation. 

\begin{figure*}[!h]
\centering
\includegraphics[width=0.48\linewidth]{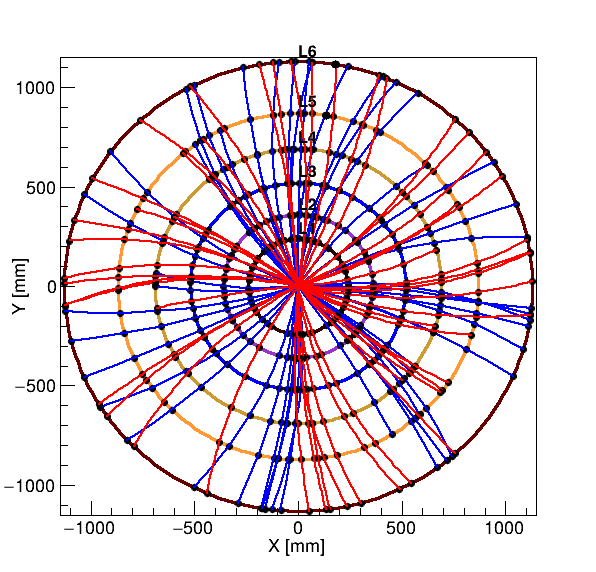}
\includegraphics[width=0.48\linewidth]{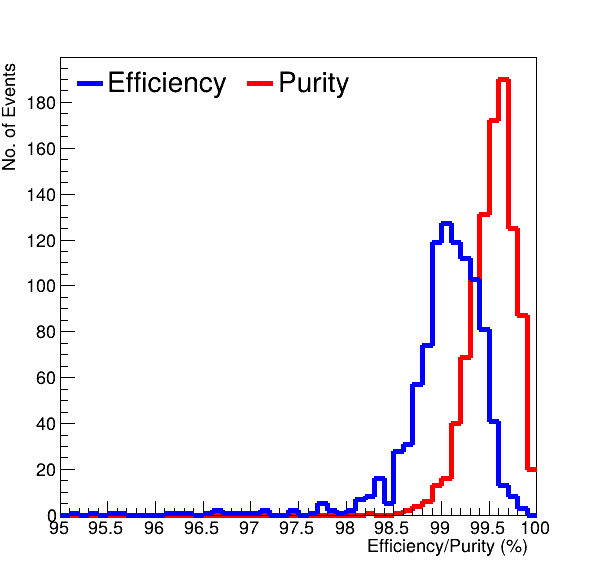}
\caption{Left: Plot of hundred tracks and tracklets (mapped \textsc{Geant4} simulated stubs) shown in transverse plane. (Blue: q = +1 and Red: q = -1). Right: Efficiency (Blue) and purity (Red) of reconstructed tracklets passing through at least four layers.}
\label{fig:geant4-eff-purity}
\end{figure*}

Clusters are identified using pixel hits. Hits with low pixel energy loss are merged with those with higher energy loss belonging to the same cluster. Association of stubs to their respective tracklets has been done using the same procedure described in Section \ref{sec:reco_pythia8}. Acceptance window on filter parameters has been applied while forming tracklets. As shown in Fig.\ref{fig:geant4-hadron-valid} (Right), the acceptance window of a given filter parameter depends moderately on the $p_t$ range. However, the $p_t$ of a track is not known at the tracklet formation stage. Therefore, the $p_t$ range is mapped to the deflection angle range between first and last (third) layer of the corresponding filter parameter. All the reconstructed tracklets satisfies the filter parameter criteria with the corresponding acceptance window. Using the true $p_t$ and azimuth angle ($\phi_0$) of each tracklet, the trajectory of the track can be drawn along with the associated stubs of the tracklet. For one of the bunch crossing events, Fig. \ref{fig:geant4-eff-purity} (Left) shows projected trajectories plotted along with its corresponding set of tracklet stubs. Reconstructed tracklets shows excellent agreement with the projected trajectories of tracks. For a better visibility only 100 tracks of an event are plotted.

\begin{figure}[!h]
\centering
\includegraphics[width=0.98\linewidth]{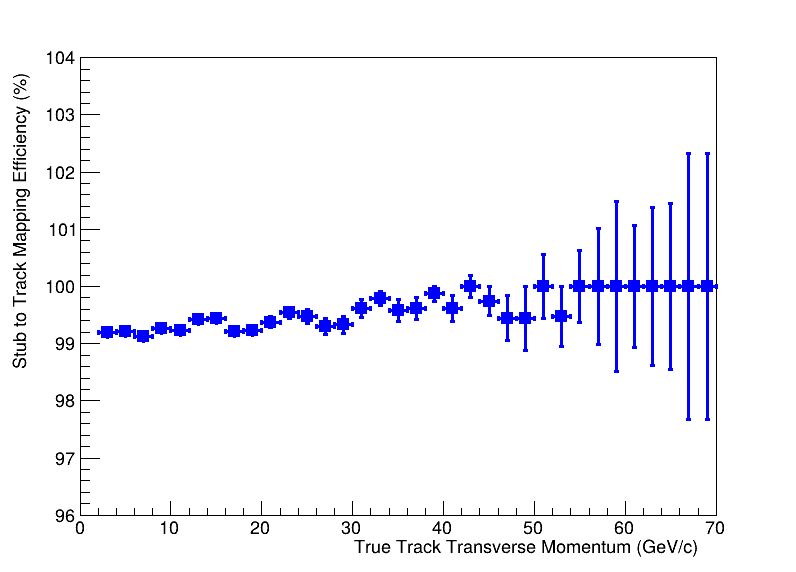}
\caption{Tracklet formation efficiency as a function of {\it true} transverse momentum of track}
\label{fig:geant4-eff-pt}
\end{figure}

The distance traveled by the particle through the detector increases with pseudorapidity, because of which energy loss as well as stub multiplicity is expected to increase. Enhancement in the stub multiplicity  may result in deviation of geometrically projected trajectory from the simulated trajectory. Hence, the efficiency (Eqn. \ref{eqn:reco_eff}) and the purity (Eqn. \ref{eqn:reco_eff}) of stubs mapping to their associated track are expected to be lower than those obtained for geometrically projected tracks. The Fig. \ref{fig:geant4-eff-purity} shows the plot of efficiency and purity for all tracks passing through at least 4 layers.
The mean efficiency is observed to be around 99\% whereas purity is more than 99.5\%. Fig. \ref{fig:geant4-eff-pt} shows the tracklet formation efficiency as a function of true transverse moment of the tracks. As can be seen from this figure, efficiency of tracklet formation does not change much with $p_t$ of the track. These figure of merits indicates excellent performance of the filter parameter based tracklet formation approach. 

\section{Track Parameter Prediction using ML-Based Techniques}
\label{sec:MLT}

The \textsc{Geant4} simulation described in Section~\ref{sec:geant4} was used to simulate 1K bunch crossing events for CMS equivalent outer tracker to be built for the HL-LHC environment. Each bunch crossing event consists of large number of pileups ($<nPU> = 150\pm12.2$) and a $t\bar t$ hard interaction resulting in about 1.9K (Fig. \ref{fig:pt_dist}) charged particles crossing at-least four layers with $p_t > 2 \ \rm GeV/c$
 and an average $p_t$ of 4.88 GeV/c. 

Collection of a large number of stubs for each event were processed using the proposed filter parameter based  algorithm (Section~\ref{sec:geant4}) to efficiently map stubs to their associated charged-particle tracks. The azimuthal coordinates of stubs in the reconstructed tracklet, $f_V \equiv (\phi_1,\phi_2,\phi_3,\phi_4,\phi_5,\phi_6)$, were used as an input feature vector for machine learning (ML) model. The {\it Decision Tree  (DT) Regressor} ML model has been used for predicting the two element target parameter vector, $T_V \equiv (q \times p_t,\phi_0^{true})$, corresponding to the charge, transverse momentum, and azimuthal angle ($\phi_0^{true}$) of the particle at the vertex. Charge is inferred from the sign of the predicted $q \times p_t$. The model performance is evaluated by comparing predicted and true kinematic parameters at the generation level. The integrated workflow, combining \textsc{Pythia8} event generation, \textsc{Geant4} detector simulation, filter parameter based tracklet formation and ML-based kinematic parameter prediction, provides an excellent {\it end-to-end} framework for charged-particle tracking studies in upcoming high-luminosity collider experiments.

\subsection{Decision Tree Regressor}
\label{sec:MLM}
A supervised learning based ML model, {\it Decision Tree  (DT) Regressor} \cite{https://doi.org/10.1002/widm.8} is used to predict the target parameter vector ($T_V$) for each reconstructed tracklet using its input feature vector ($f_V$). The model is trained using truth-level kinematic information on $f_V$ and $T_V$ obtained from the generator level data. The Decision Tree (DT) regressor is employed due to its relative conceptual simplicity compared with more complex ensemble and deep-learning models, and its ability to capture nonlinear relationships between the input features and target track parameters. The model recursively partitions the feature space by greedily selecting, at each node, the feature and threshold resulting in maximum reduction in weighted squared-error. Since DT models do not require explicit feature scaling, they are well suited for handling reconstructed tracklet features directly. The DT regressor is trained using $Scikit-learn$ \cite{pedregosa2018scikitlearnmachinelearningpython} framework to predict the kinematic parameters of charged-particle tracks.

\subsection{ML Model Training Strategy}

\begin{figure}[!h]
    \centering
    \includegraphics[trim=2pt 2pt 2pt 2pt, clip,width=0.98\linewidth]{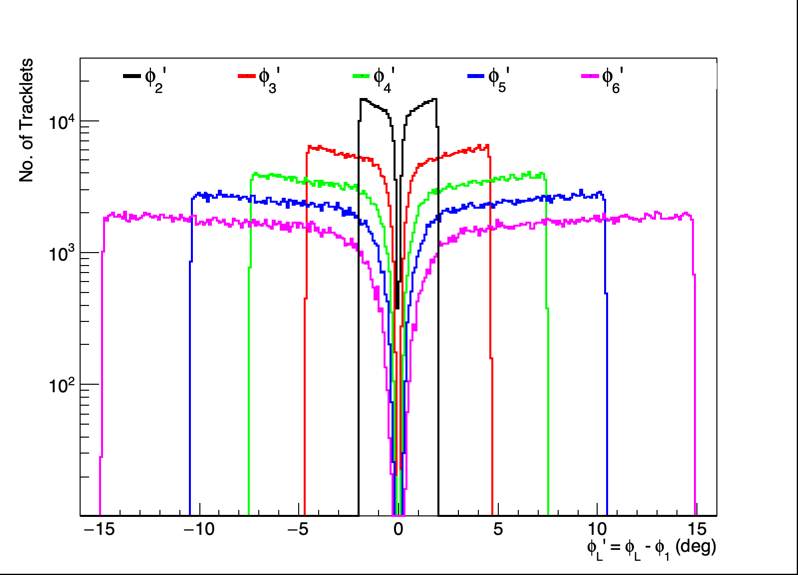}
    \caption{ Distributions of input features ($\Phi'_{2}...\Phi'_{6}$) used in the ML model.}
    \label{fig:inputfeature}
\end{figure}

Training of the DT model described in Section~\ref{sec:MLM} follows a structured strategy using a common set of input feature vectors ($f_V$) derived from the reconstructed tracklets. Rotational symmetry of the input feature is used to reduce the redundancy in the parameter space of feature vector and emphasize on the relative information without loss of any generality. It makes the training process efficient and accurate. The input features are transformed by subtracting the azimuth of the first tracker layer, {\it i.e.}, $\phi'_L = \phi_L - \phi_1$ (L=1,2,...6). The transformation results in a modified input feature vector $f_V' = (\phi'_2, \ \phi'_3, \ \phi'_4, \ \phi'_5, \ \phi'_6)$  which is used for the training as well as prediction. It is to be noted that $\phi_1'$, by definition, is identically zero for all modified input feature vectors. Elements of $f_V'$ represents the deflection angle {\it w.r.t.} the first layer and their respective distribution in the training sample are shown in Fig.~\ref{fig:inputfeature}. The deflection angle is inversely proportional to the transverse momentum and it increases with layer number. The distribution falls rapidly towards low deflection angle due to exponentially falling $p_t$ distribution of the generated particles. Supervised training is performed using the truth-level kinematic information ($q \times    p_t, \ \phi_0'$) of each charged particle, where, $\phi_0' = \phi_0^{true}-\phi_1$. The training for each component of the target parameter vector,  $\phi_0'$ and $q \times p_t$, is carried out separately. Due to the falling spectrum of $p_t$, statistics for high $p_t$ particles are significantly smaller than particles with low $p_t$. Also, the relationship between $f_V'$ and the target parameter component $p_t$ is highly nonlinear and may vary significantly across different momentum regions. In view of this, the DT model employs a binned $p_t$ training strategy. It helps in reducing the statistical imbalance of the input data and also enables models in broadly classifying the varying detector response and track curvature behavior across different momentum ranges. 

The simulated dataset consists of 1K independent bunch crossing events, each composed of hard collision and large number of pile-up collisions. 70\% of the dataset is used for training the model, whereas the remaining 30\% events are exclusively used for the model evaluation. Accordingly, all reconstructed tracklets belonging to a given bunch crossing are assigned entirely to either the training or the evaluation dataset, ensuring an unbiased assessment of the model performance. In addition to the inclusive ($|q \times p_t| \geq 2~\mathrm{GeV}/c$) training dataset, the same dataset is further divided into several overlapping $|q \times p_t|$ intervals: $[2,6], \ [5,10], \ [9,15], \ [14,20] $ and $ \ >19 \ \rm GeV/c$. Separate models are trained for each interval to improve the prediction accuracy across the full momentum range. For the second target parameter ($\phi_0'$), training is performed using the entire training dataset without any $p_t$ binning, as the azimuth angle ($\phi_0^{true}$) of particles at vertex is uniformly distributed.

\begin{figure}[!h]
\centering
\includegraphics[width=0.98\linewidth]{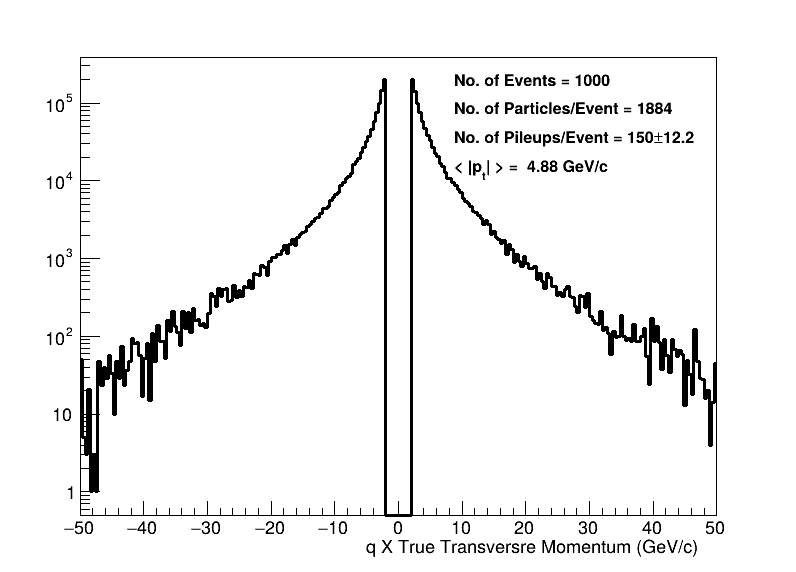}
\caption{True transverse momentum distribution of particles ($q=\pm1$) passing through at-least four layers}
\label{fig:pt_dist}
\end{figure}

\subsection{Evaluation Strategy}
The performance of the trained ML model is evaluated using the independent reconstructed tracklets that were not used for training. For each reconstructed tracklet, its corresponding feature vector ($f_V'$) is fed to the trained models to predict the parameters of the target vector ($T_V'$). A two-stage evaluation strategy is adopted for the prediction of $q \times p_t$. First, an inclusive DT model trained over the entire $q \times p_t$ range is used to get the initial estimate of $q \times p_t$. Based on the initial estimate, the same tracklet feature vector ($f_V'$) is fed to the corresponding $p_t$ binned DT model, using which the final prediction of $q \times p_t$ is obtained. It is to be noted that due to overlap in the $p_t$ binning, if the predicted $q \times p_t$ corresponds to two adjacent $p_t$ binned DT models, then the model with the higher $p_t$ range is chosen to get the final target parameter $q \times p_t$. Such strategy is adopted due to the relatively higher statistical strength in the higher $p_t$ bin data. Two-stage prediction of the target parameter $q \times p_t$ improves the momentum resolution in the entire $p_t$ region. Prediction of $\phi_0'$ is performed using a single-stage evaluation approach by feeding the same feature vector $f_V'$ to the model. The final target parameter $\phi_0^{pred}$ is obtained by adding the offset ($\phi_0^{pred} = \phi_0' + \phi_1$). 

\begin{figure}[!h]
\centering
\includegraphics[trim=2pt 2pt 2pt 2pt, clip,width=0.98\linewidth]{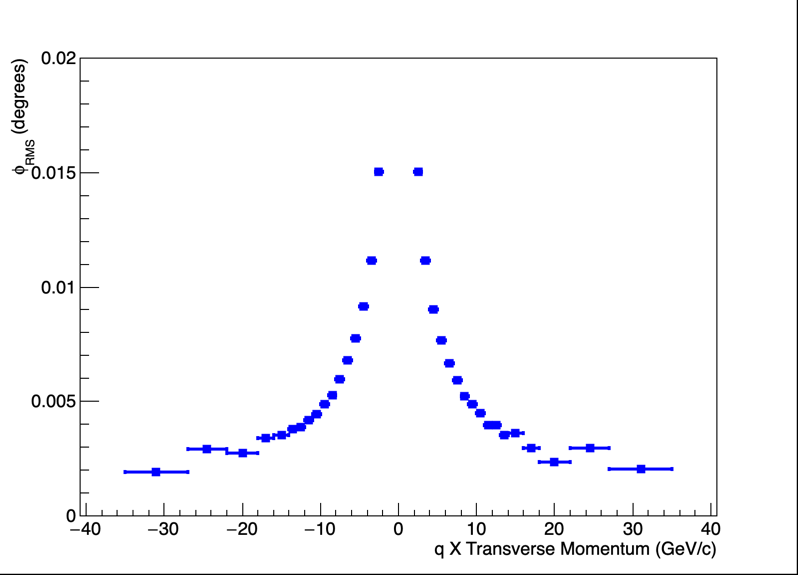}
\caption{Residual azimuth ($\phi_0^{true}-\phi_{0}^{pred}$) at vertex as a function of transverse momentum of particles passing through at-least four layers}
\label{fig:phirms}
\end{figure}

\begin{figure*}[!h]
\centering
\includegraphics[trim=2pt 2pt 2pt 2pt, clip,width=0.48\linewidth]{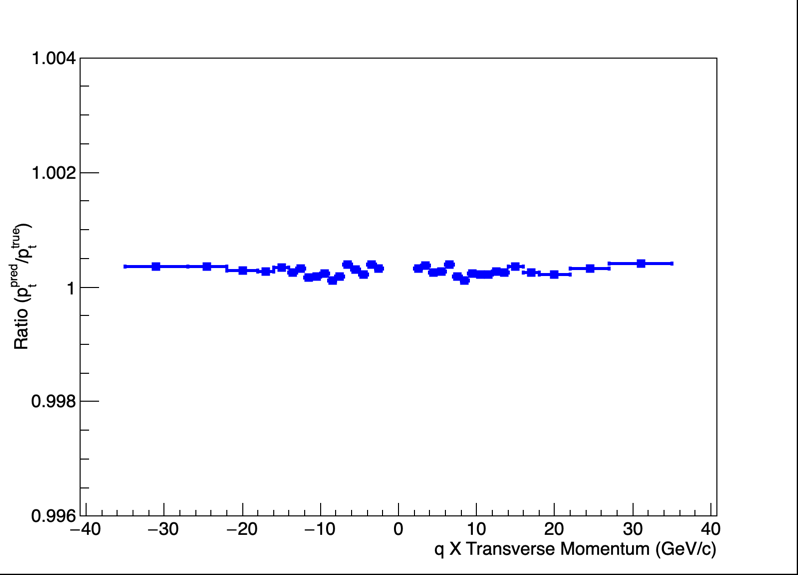}
\includegraphics[trim=2pt 2pt 2pt 2pt, clip,width=0.48\linewidth]{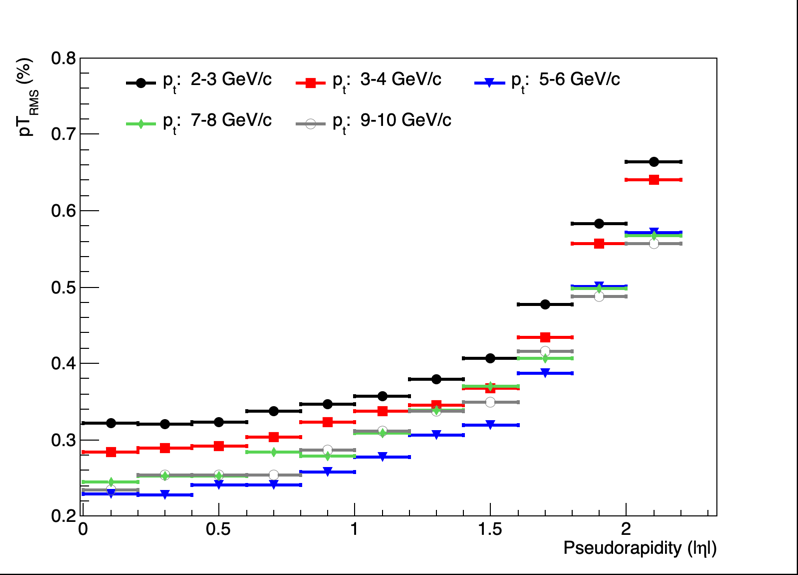}
\caption{Left: Ratio of predicted and true $p_t$ as a function of true transverse momentum. Right: Resolution of transverse momentum predicted using DT model as a  function pseudorapidity in several $p_t$ ranges. Tracks are required to pass through at least 4 layers.}
\label{fig:pt_ratio_resln}
\end{figure*}

Performance of the predicted $\phi_0$ at vertex is examined by obtaining the distribution of residual azimuth, defined as $\Delta \phi_0 = \phi_0 - \phi_0^{true}$ in different bins of transverse momentum. The plot of azimuth angle resolution ($\phi_{RMS}$) at the vertex as a function of $q \times p_t$ is shown in Fig.~\ref{fig:phirms}. The plot  demonstrates the excellent agreement between predicted and true azimuth angle of a particle at vertex across the entire momentum range of particles. At lower $p_t$, the azimuth resolution is about $0.015^\circ$  which flattens around $0.003^\circ$ at higher $p_t$. Performance of another important target parameter, $q \times p_t$, is probed  by studying the ratio, $R = p_t^{pred}/p_t^{true}$ for each particle. The mean and RMS ($pT_{RMS}$) of the ratio for each of the distributions is obtained as a function of the pseudorapidity in each $p_t$ range. The mean of the ratio closer to the unity reflects linearity of the response. As shown in Fig.~\ref{fig:pt_ratio_resln} (Left) the ratio of the predicted and true $p_t$ of the particles, integrated over $\eta$, is very close to the unity, indicating excellent linearity of the response over a wide range of transverse momentum of the particles. 
Resolution of the transverse momentum as a function of $\eta$ in different $p_t$ ranges is shown in the Fig.~\ref{fig:pt_ratio_resln} (Right). The pseudorapidity region $|\eta|<1.8$, $1.8< |\eta|<2.0$ and $2.0<|\eta|<2.2$ are dominated by tracks exclusively passing through 6, 5 and 4 layers respectively. Therefore, resolution worsens in higher $\eta$ region ($|\eta|>1.8$) due to the lower number of sampling points. The $pT_{RMS}$ value in $|\eta|<1.8$ region moderately increases with $\eta$. Overall resolution in this region varies between $0.22\%$ to $0.42\%$ in the $p_t$ range of 2-10 GeV/c. The $\eta$ dependence of momentum resolution at higher $p_t$ could not be obtained due to statistics limited sample. However overall $p_t$ resolution remains in the same region for $|q \times p_t|<35 \ GeV/c$.

\section{Summary and Discussion}

In this work, we present a novel track reconstruction algorithm designed for the Phase-2 tracker of the CMS Experiment to be operated during high luminosity era.  The proposed approach introduces a new strategy for formation of tracklets which is fundamentally different than the conventional approaches used in the pp collider experiments. The central feature of this method is the proposed filter parameter ($\alpha$), which serves as a powerful discriminant for precise and direct formation of tracklets, i.e., mapping stubs to their corresponding particle trajectory in the tracker. The filter parameters helps in efficiently eliminating fake combinatorics at the early stage of tracklet formation. The filter parameters ($\alpha$) are defined for every three consecutive layers, hence there are  four filter parameters, namely, $\alpha_{13}, \ \alpha_{24}$, $\alpha_{35}$ and $\alpha_{46}$, corresponding to the six-layer tracker geometry considered in this study. These parameters encode the track bending information induced by the CMS solenoidal magnetic field of 3.8~T. Rotational symmetry in the azimuthal plane is used to eliminate the $\phi$ dependent divergences in the filter parameters (Fig. \ref{fig:alpha}) estimated using azimuth coordinates of pixel hits (Eqn. \ref{eqn:alpha}) from three consecutive layers of a potential tracklet candidate. 

Performance of the proposed algorithm was evaluated by first testing it on the generator-level particle information obtained from simulated $t\bar{t}$ events in high pile-up ($nPU=150\pm12.2$) conditions using \textsc{Pythia8}. Such a scenario is representative of the expected conditions at the HL-LHC. Each generated particle using its kinematic parameters ($\eta_0, z_0, \phi_0$ and $p_t$) was {\it geometrically projected} into the tracker to get the  accurate coordinates of its intersection with each layer. Proposed algorithm was applied on collection of hits of all particles belonging to the same bunch crossing event to build tracklets. Detailed  study performed on reconstructed tracklets shows excellent efficiency and high purity (Fig. \ref{fig:pythia8_tracks}) with the proposed filter parameter based algorithm. 

In order to incorporate realistic detector effects, a simplified tracker geometry using \textsc{Geant4} was developed. The design was kept similar to the CMS outer tracker design for HL-LHC. Particle trajectory simulation incorporates effects of pixel resolution, magnetic field, multiple hits in same layer, energy loss mechanism {\it etc.} allowing to evaluate the robustness of the algorithm under more realistic experimental conditions. All particles simulated using \textsc{Pythia-8} for each bunch crossing were passed through the detector geometry using \textsc{Geant4} framework. Our novel reconstruction method was applied to reconstruct tracklets using large number of simulated hits. Excellent efficiency and purity across the entire $p_t$  range (Fig. \ref{fig:geant4-eff-purity} and \ref{fig:geant4-eff-pt}) demonstrates strong performance of the algorithm to efficiently reconstruct tracklets under high luminosity phase of the LHC.

A pipeline for reconstruction of transverse momentum ($p_t$) and azimuthal angle ($\phi_0$) at vertex of all particles has been implemented using a machine learning based regression approach. The decision tree model was deployed to predict $q\times p_t$ and $\phi_0$ independently. The model was trained with tracklets built using proposed algorithm. Due to steeply falling $p_t$ spectrum of particles (Fig. \ref{fig:pt_dist}), issue of statistical imbalance in the training data was mitigated by splitting the data in several $p_t$ bins. Training and evaluation of each dataset was done independently. A two-tier approach for predicting $p_t$ has yielded higher accuracy of the prediction. The azimuthal angle ($\phi_0$) data is uniformly distributed across the entire range, thus the prediction of $\phi_0$ was carried out in a single step. Excellent $p_t$ and $\phi_0$ resolution has been obtained using this approach. The $p_t$ resolution varies between  0.2\% to 0.7\% for a $p_t$ and $\eta$ range of $2<p_t<35 \ \rm GeV/c$ and $|\eta|<2.2$ respectively (Fig. \ref{fig:pt_ratio_resln}). The $p_t$ resolution has a moderate dependence on $\eta$ and drops at higher pseudorapidity ($|\eta|>1.8$) due to limited detector acceptance resulting in reduction in number of sampling points. The $\phi_0$ resolution varies between $0.002^\circ - 0.016^\circ$ across entire $p_t$ and $\eta$ range (Fig. \ref{fig:phirms}). The efficiency, purity and resolution has been seen to be comparable with expected performance of the Phase-2 CMS tracker ~\cite{CERN-LHCC-2017-009} thus validating the entire pipeline primarily based on a new filter parameter based tracklets formation and machine learning model. 

Proposed filter parameter based algorithm provides a promising alternative strategy for track reconstruction in the high-occupancy environment expected at the HL-LHC. Same pipeline can be further extended to build tracklets emerging from long lived neutral particles decaying into charged particles thereby enhancing the potential to probe beyond standard model phenomenon. The algorithm has an inherent potential for exploiting parallel processing architectures seamlessly to significantly enhance the computing performance. Embedded systems can also be explored to encapsulate this pipeline.

\begin{acknowledgements}
The authors gratefully acknowledge the High Performance Computing (HPC) facility at the Indian Institute of Science Education and Research (IISER) Mohali for providing the computational resources used in this work.
\end{acknowledgements}


\bibliographystyle{spphys}
\bibliography{references}

\end{document}